\documentclass[mathleft]{an}
\usepackage{amsmath}
\usepackage{amssymb}
\usepackage{textcomp}
\usepackage{graphicx}
\usepackage{times}
\usepackage{changes}
\usepackage{supertabular}
\begin{document}
\Pagespan{1}{}
\Yearpublication{--}
\Yearsubmission{--}
\Month{--}
\Volume{-}
\Issue{-}

\title{Spectroscopic characterization of nine binary star systems as well as HIP\,107136 and HIP\,107533 \thanks{Based on observations obtained with telescopes of the University Observatory Jena, which is operated by the Astrophysical Institute of the Friedrich-Schiller University.}}

\author{T.~Heyne\fnmsep\thanks{Corresponding author: T.~Heyne, Astrophysikalisches Institut und Universit\"{a}ts-Sternwarte Jena, Schillerg\"{a}{\ss}chen 2-3, 07745 Jena, Germany. \email{therese.heyne@uni-jena.de}\newline}, M.~Mugrauer, R.~Bischoff, D.~Wagner, S.~Hoffmann, O.~Lux, V.~Munz, M.~Geymeier, R.~Neuh\"{a}user}

\titlerunning{Spectroscopic characterization}
\authorrunning{Heyne et al.}
\institute{Astrophysikalisches Institut und Universit\"{a}ts-Sternwarte Jena, Schillerg\"{a}{\ss}chen 2-3, 07745 Jena, Germany}

\received{--}
\accepted{--}
\publonline{later}

\keywords{binaries: spectroscopic, stars: individual: HIP\,107162, HIP\,23040, HIP\,2225, HIP\,30247, HIP\,113048, HIP\,25048, HIP\,85829, HIP\,77986, HIP\,98194, HIP\,107136, HIP\,107533}

\abstract{We present the results of our 2nd radial velocity monitoring campaign, carried out with the \'Echelle spectrograph FLECHAS at the University Observatory Jena in the course of the Gro{\ss}schwabhausen binary survey between December 2016 and June 2018. The aim of this project is to obtain precise radial velocity measurements for spectroscopic binary stars in order to redetermine, verify, improve and constrain their Keplerian orbital solutions. In this paper we describe the observations, data reduction and analysis and present the results of this project. In total, we have taken 721 RV measurements of 11 stars and derived well determined orbital solutions for 9 systems (7 single-, and 2 double-lined spectroscopic binaries) with periods in the range between 2 and 70\,days. In addition, we could rule out the orbital solutions for the previously classified spectroscopic binary systems HIP\,107136 and HIP\,107533, whose radial velocities are found to be constant on the km/s-level over a span of time of more than 500\,days. In the case of HIP\,2225 a significant change of its systematic velocity is detected between our individual observing epochs, indicating the presence of an additional companion, which is located on a wider orbit in this system.}
\maketitle

\section{Introduction}

The University Observatory Jena is located close to the village Gro{\ss}schwabhausen roughly 10\,km west of the city Jena (\cite{pfau}). Since 2013 the fibre-linked \'{E}chelle spectrograph FLECHAS (\cite{mugrauer2014}) is mounted at the 90-cm reflector telescope while it is operated in the Nasmyth mode ($D=90$\,cm, $f/D=15$). As part of the Gro{\ss}schwabhausen binary survey (\cite{mugrauer2016}) a radial velocity (RV) monitoring program was started in 2015 (\cite{bischoff}) in order to redetermine, verify, improve and constrain the Keplerian orbital solutions of spectroscopic binaries, selected from the 9th Catalogue of Spectroscopic Binary Orbits (SB9 hereafter, see \cite{pourbaix2004}). The SB9 yields information about orbital elements of several thousand spectroscopic binary systems. The binaries are ranked by different orbit grades with values from 0 (poor) to 5 (definitive), dependant on the quality of their orbital solutions.

The project presented here is a continuation of this RV monitoring programm with a target sample, chosen from the SB9, with less constrained orbital solutions, which exhibit orbital grades of 3 or lower. The properties of the selected targets are summarized in Tab.\,\ref{table_targetproperties}. Section 2 of this paper describes in detail the spectroscopic observations, taken in this project, and the data reduction. Section 3 provides information about the procedure of measuring individual radial velocities. The derived Keplerian orbital solutions are presented in the Section 4 and their properties are discussed in the last section of this paper.

\begin{table*}[h!]
\centering \caption{Properties of the observed targets. For all targets their equatorial coordinates (RA, Dec), V-band magnitude ($V$) and
spectral type (SpT) are given together with their orbital solution (period $P$, periastron time $T_0$, eccentricity $e$, argument of periastron $\omega$, RV semi-amplitude $K$, and systematic velocity $\gamma$) as listed in the SB9 with an orbit grade of 3 or lower.}
\resizebox{1\linewidth}{!}{
\begin{tabular}{ccccccccccccc}
	\hline
	Target & RA$_{\text{J2000}}$ & Dec$_{\text{J2000}}$ & $V$ & SpT & Grade & $P$ & $T_0$ & $e$ & $\omega$ & $K$ & $\gamma$ & Ref. \\
&  $[$hh mm ss.ss$]$ & $[$dd mm ss.s$]$ & $[$mag$]$ & & & $[$d$]$ & $[JD]$ & &  $[^\circ]$ & $[$km/s$]$ & $[$km/s$]$\\
	\hline
HIP\,107162 & 21 42 22.96	& $+$41 04 37.3	& 5.74 & A0V      &	2     & 1.729   & 2424255.8   & 0.0  & 0.0   & 110$^{*}$ & $-$25.5 & $[$a$]$\\
HIP\,23040  & 04 57 17.20	& $+$53 45 07.6	& 4.43 & A1V      & 3     & 3.8845  & 2418686.7   & 0.0  & 0.0   & 35.8      & $-$9.5  & $[$b$]$\\
HIP\,2225   & 00 28 13.65	& $+$44 23 40.0	& 5.20 & A2Vs     & 3     & 3.9558  & 2418841.6   & 0.15 & 233.2 & 41.7      & $+$2.0  & $[$c$]$\\
HIP\,30247  & 06 21 46.13	& $+$53 27 07.8	& 5.36 & F5III    & 3     & 6.5013  & 2423634.2   & 0.02 & 330.6 & 31.7      & $-$1.5  & $[$d$]$\\
HIP\,113048 & 22 53 40.16	& $+$44 44 57.0	& 5.81 & A3m      & 1     & 24.1635 & 2440000.4   & 0.2  & 5.0   & 10.1      & $+$6.7  & $[$e$]$\\
HIP\,25048  & 05 21 48.42	& $+$41 48 16.5	& 5.22 & B5V      & 1     & 35.5	& 2430000.0   & 0.0  & 0.0   & 28.0      & $+$19.0 & $[$f$]$\\
HIP\,85829  & 17 32 16.03	& $+$55 10 22.6	& 4.89 & A4m      & 2     & 38.128  & 2440022.4   & 0.04 & 0.6   & 9.8       & $-$16.7 & $[$e$]$\\
HIP\,77986  & 15 55 30.59	& $+$42 33 58.3	& 5.75 & B7IV-Ve  & 1     & 46.194  & 2441473.8   & 0.34 & 5.0   & 12.0      & $-$19.3 & $[$g$]$\\
HIP\,98194  & 19 57 13.87	& $+$40 22 04.2	& 5.41 & B5Vp     & 2     & 70.22   & 2439762.1   & 0.34 & 302   & 41.4      & $-$20.1 & $[$h$]$\\
HIP\,107136 & 21 42 05.67	& $+$51 11 22.6	& 4.67 & B3V      & 1     & 26.33	& 2431306.5   & 0.0  & 0.0   & 16.5      & $-$8.2  & $[$i$]$\\
HIP\,107533	& 21 46 47.61	& $+$49 18 34.5 & 4.24 & B3III    & 2     & 72.0162 & 2428410.6   & 0.34 & 238.1 & 7.8       & $-$12.3 & $[$j$]$\\
\hline
\end{tabular}}\\
\flushleft
$^*$ Amplitude of both components of this double-lined spectroscopic binary system.\\ \vspace{0.1cm}
~~~References for the listed orbital solutions from the SB9:\\ \vspace{0.1cm}
\begin{tabular}{lllll}
$[$a$]$ \cite{luyten} & $[$b$]$ \cite{lucy}  & $[$c$]$ \cite{udick}  & $[$d$]$ \cite{harper} & $[$e$]$ \cite{abt}\\
$[$f$]$ \cite{blaauw} & $[$g$]$ \cite{heard} & $[$h$]$ \cite{batten} & $[$i$]$ \cite{fehrenbach} & $[$j$]$ \cite{taffara}\\
\end{tabular}
\rule{\textwidth}{0.5pt}
\label{table_targetproperties}                                              		
\end{table*}

\section{Observations and data reduction}

In the course of our RV monitoring project 11 targets were observed during two observing epochs between December 2016 and May 2017, and from January to June 2018. In total, more than 33 spectra could be taken for each target with FLECHAS in its 1x1 binning mode, which provides a spectral resolving power of $R \sim 9300$. On average, the obtained spectra of all targets exhibit a signal-to-noise-ratio ($SNR$) in the range between about 110 and 300 (as measured at a wavelength of 6500\,\AA), which is adequately high to obtain accurate RV measurements. In general three spectra, each with a minimum integration time of 150\,s were taken for each target to remove cosmics from the resulting spectra, and to achieve a sufficiently high $SNR$. Directly before the spectroscopy of each target 3 flat-field frames of a tungsten lamp with an exposure time of 5\,s were taken followed by 3 spectra of a ThAr lamp (with more than 700 detected emission lines) for wavelength calibration.

In addition, dark frames for all used exposure times were taken in each observing night for the dark current removal. The complete spectroscopic data reduction was done with the FLECHAS software pipeline, developed at the Astrophysical Institute Jena, which is optimized for the reduction of FLECHAS data. The software includes dark current subtraction, flat-fielding, extraction and wavelength calibration of the individual spectral orders, as well as the averaging and normalization of all spectroscopic data (\cite{mugrauer2014}). Throughout this project, 2163 spectra of all targets could be taken, which results in 721 fully reduced spectra with a total integration time of 132\,h. The details of the observations of all targets are summarized in the observation log, which is listed in Tab.\,\ref{table_log}.

\begin{table}[h!]
\centering\caption{Observation log. For each target we list the number of observations ($N_{\text{Obs}}$), the dates of the first and last successful observation, as well as the average signal-to-noise ratio ($<SNR>$) of all spectra, as measured at $\lambda = 6500$\,\AA.}
\resizebox{1\linewidth}{!}{
\begin{tabular}{ccccc}
\hline
Target            & $N_{\text{Obs}}$   & first obs. & last obs.       & $<SNR>$ \\
\hline
HIP\,107162	      & 53                 & 2016 Dec 5 & 2018 Jun 3      & 117 \\
HIP\,23040	      & 104                & 2016 Dec 6 & ~~~~2018 May 25 & 167 \\
HIP\,2225	      & 36                 & 2016 Dec 6 & 2018 Jun 3      & 122 \\
HIP\,30247	      & 111                & 2016 Dec 6 & ~~~~2018 May 28 & 134 \\
HIP\,113048	      & 42                 & 2016 Dec 5 & 2018 Jun 3      & 119 \\
HIP\,25048	      & 69                 & 2016 Dec 5 & ~~2018 May 1    & 114 \\
HIP\,85829	      & 77                 & 2016 Dec 5 & 2018 Jun 3      & 162 \\
HIP\,77986	      & 63                 & 2016 Dec 5 & 2018 Jun 3      & 117 \\
HIP\,98194	      & 59                 & 2016 Dec 5 & 2018 Jun 3      & 147 \\
HIP\,107136	      & 55                 & 2016 Dec 5 & 2018 Jun 3      & 141 \\
HIP\,107533	      & 52                 & 2016 Dec 5 & 2018 Jun 3      & 193 \\
\hline \\
\end{tabular}}
\label{table_log}                                              		
\end{table}

\section{Radial velocity measurements}

\begin{figure*}
 \includegraphics[width=1\textwidth]{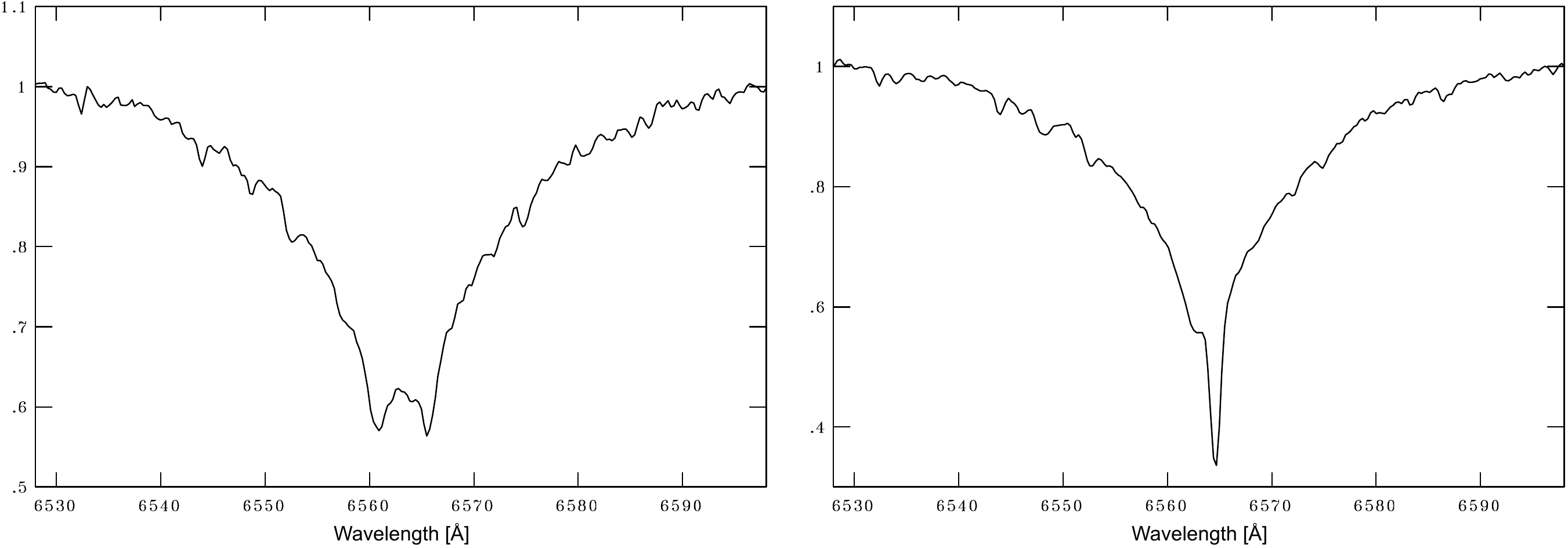}
  	\caption{The H$_{\alpha}$-line profile of HIP\,107162 (left) and HIP\,2225 (right), extracted from the FLECHAS spectra of the stars taken on 2016 Dec 21. The SB2 nature of both stars is clearly detected. While the spectral lines of the individual components of HIP\,107162 are resolved in the FLECHAS spectra, taken at orbital phases with maximal RV deviations from the systematic velocity, the lines of the components of HIP\,2225 remain significantly blended at all orbital phases.}
 \label{fig:sb2}
\end{figure*}

The RVs of the targets are determined for all observing dates in the fully reduced spectra of the stars, by measuring the wavelengths of the first three Hydrogen Balmer lines (H$_\alpha$: $\lambda_0 =  6562.81$\,\AA, H$_\beta$: $\lambda_0 = 4861.34$\,\AA, H$_\gamma$: $\lambda_0 = 4340.48$\,\AA), which are prominent lines in the spectra of early type stars with spectral type B to F. Thereby, the wavelength of the spectral lines are determined by fitting a Gaussian profile on the Doppler broadening dominated line cores, using the IRAF standard script \textsc{splot}.
The RV of the stars is derived from the measured Doppler shifts ($\lambda - \lambda_{0}$) of the individual spectral lines by

\begin{equation}
RV = c \cdot \frac{\lambda - \lambda_{0}}{\lambda_{0}} + BC
\end{equation}

with $c$ the speed of light, and $BC$ the barycentric correction.

In the fully reduced spectra the RVs of all stars, obtained by the core fitting of their Balmer lines, are averaged and their standard deviation is determined. This yields for all observing dates the RV of each target, including its uncertainty.

In addition, telluric spectral lines, detected in the different spectral orders are used to check the stability of the wavelength calibration of FLECHAS throughout our RV monitoring project. The RVs of the measured telluric lines scatter on average on the $0.9\, \text{km}\,\text{s}^{-1}$ level, which is consistent with the reached RV precision of the different observed targets. The long-term RV stability of the spectrograph was already verified by \cite{irrgang} and \cite{bischoff}, and is also confirmed by our measurements, taken in the course of this project.

While most of the observed targets are classified as single-lined spectroscopic binaries, a splitting of the spectral lines is detected in some spectra of HIP\,107162, and HIP\,2225, as it is illustrated in Fig.\,\ref{fig:sb2}. Hence, these stars are double-lined spectroscopic binaries, in which the RVs of their components can be measured separately.

\begin{figure}
\resizebox{\columnwidth}{!}{\includegraphics{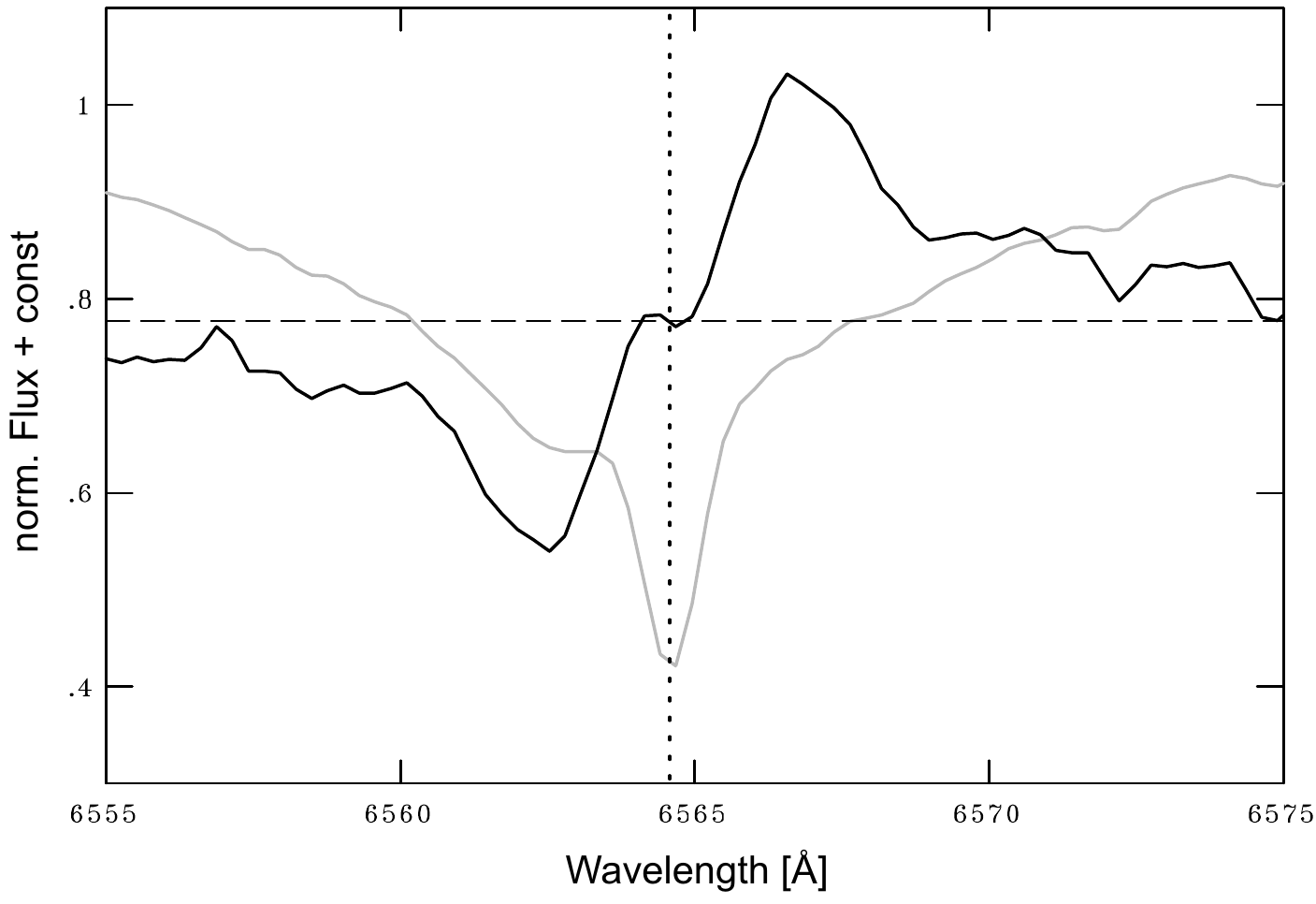}}
  	\caption{Detail view of the H$_{\alpha}$ line-profile (grey line, with its center indicated with a dotted vertical line) of HIP\,2225, extracted from the normalized FLECHAS spectrum of the star, taken on 2016 Dec 21. The difference spectrum, in which the spectral line of the primary component is subtracted, is illustrated as dark line with its zero normalized flux level, indicated as dashed line. The spectral line of the fainter secondary component clearly appears as positve/negative pair in the difference spectrum.}
 \label{fig2:sb2}
\end{figure}

In order to obtain the wavelengths of the blended spectral lines of the observed SB2s the line-deblending task in \textsc{splot} is used if applicable. In the case of HIP\,107162 the wavelengths of the spectral lines are determined using line-deblending at orbital phases with maximal RV deviations from the systematic velocity. In contrast for HIP\,2225 the spectral lines of the individual components remain significantly blended even at extreme orbital phases, and the line-deblending fails to provide the wavelengths of both components, only the wavelength of the spectral lines of the primary star can be measured. In order to determine the wavelengths of the spectral lines of the secondary component we flip all normalized spectra at the line centers of the primary and subtract them from the original spectra. As illustrated in Fig.\,\ref{fig2:sb2}, in these difference spectra the line profile of the primary component is well removed and only the spectral line of the secondary component remains (as a positive/negative pair), whose wavelength is determined by Gaussian fitting.

The determined RVs of our targets with their uncertainties are summarized for all observing dates (given as barycentric Julian date $BJD$) in Tab.\,\ref{tab:HIP107162} to \ref{tab:HIP107533}. The reached RV precision ranges between 0.6 and 2.8\,km/s for the single-lined, and between 2.3 and 20.4\,km/s for the double-lined spectroscopic binaries, respectively.

\section{Orbit determination}

\begin{figure*}
\centering{\includegraphics[height=0.99\textheight]{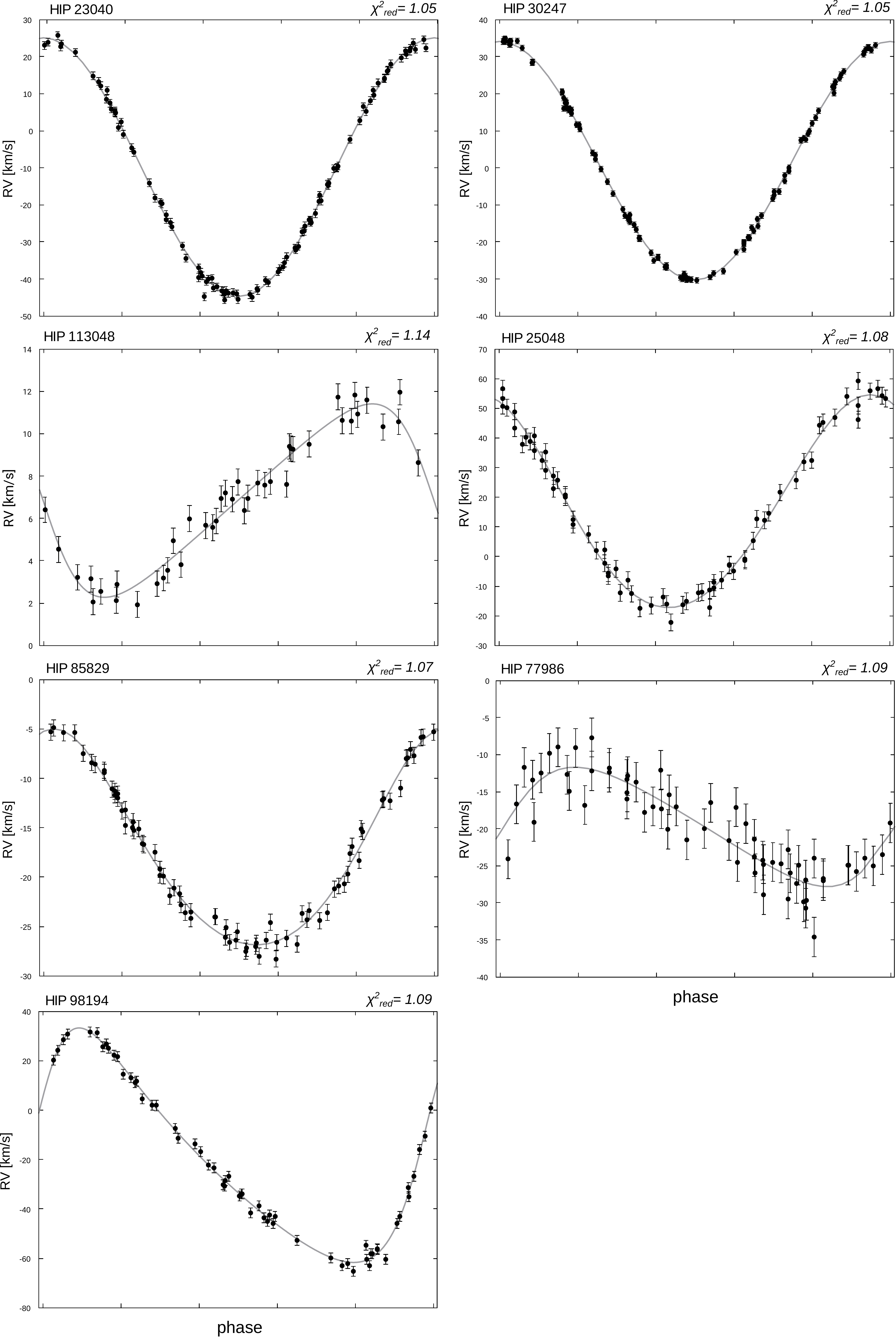}}
 	\caption{The phase-folded RV curves of the 7 single-lined spectroscopic binaries, observed in the course of our project. For all targets their RV measurements are shown with their uncertainties together with the derived best fitting Keplerian orbital solutions (grey lines).}
 \label{fig:elements}
\end{figure*}

\begin{table*}[h!]
	\caption{The derived Keplerian orbital elements with their uncertainties of all spectroscopic binaries, observed in this project.}
	\centering
	\resizebox{1\linewidth}{!}{
	\begin{tabular}{lcccccc}
		\hline
\multicolumn{1}{c}{Target} & \multicolumn{1}{c}{$P$[d]} & \multicolumn{1}{c}{$T$ [$BJD$]} & \multicolumn{1}{c}{$e$} & \multicolumn{1}{c}{$\omega$ [$^\circ$]}  & \multicolumn{1}{c}{$K$ [km/s]} & \multicolumn{1}{c}{$\gamma$ [km/s]} \\
\hline
HIP\,107162\,A	 & ~~$1.728 \pm 0.061$  & $2457726.95 \pm 0.01$ & $0$ [fixed]       & $180$ [fixed]        & ~~$96.2 \pm 3.9$  & $-16.3 \pm 2.0$~~ \\
HIP\,107162\,B	 & ~~$1.728 \pm 0.061$  & $2457726.95 \pm 0.01$ & $0$ [fixed]       & $0$ [fixed]          & $107.4 \pm 3.9$   & $-16.3 \pm 2.0$~~ \\
HIP\,23040	     & ~~$3.884 \pm 0.001$  & $2457726.05 \pm 0.01$ & $0$ [fixed]       & $0$ [fixed]          & ~~$34.9 \pm 0.2$  & $-9.8 \pm 0.1$    \\
HIP\,2225\,A	 & ~~$3.956 \pm 0.001$  & $2457727.19 \pm 0.06$ & $0.136 \pm 0.011$ & $276.0 \pm 5.4$~~    & ~~$48.4 \pm 1.0$  & $0$ [fixed]       \\
HIP\,2225\,B	 & ~~$3.956 \pm 0.001$  & $2457727.19 \pm 0.06$ & $0.136 \pm 0.011$ & $ 96.0 \pm 5.4$      & ~~$86.5 \pm 1.0$  & $0$ [fixed]       \\
HIP\,30247	     & ~~$6.501 \pm 0.001$  & $2457726.71 \pm 0.01$ & $0$ [fixed]       & $0$ [fixed]          & ~~$32.2 \pm 0.1$  & $+1.9 \pm 0.1$    \\
HIP\,113048	     & $24.190 \pm 0.022$   & $2457724.80 \pm 0.40$ & $0.296 \pm 0.035$ & $90.6 \pm 8.9$       & ~~~~$4.6 \pm 0.2$ & $+6.9 \pm 0.1$    \\
HIP\,25048	     & $34.507 \pm 0.013$   & $2457725.50 \pm 1.00$ & $0.073 \pm 0.014$ & ~~$20.5 \pm 10.9 $   & ~~$35.9 \pm 0.5$  & $+16.2 \pm 0.4$~~ \\
HIP\,85829	     & $38.058 \pm 0.013$   & $2457714.88 \pm 0.66$ & $0.128 \pm 0.012$ & $347.4 \pm 6.1$~~    & ~~$10.9 \pm 0.1$  & $-17.3 \pm 0.1$~~ \\
HIP\,77986	     & $45.862 \pm 0.108$   & $2457723.93 \pm 2.26$ & $0.234 \pm 0.059$ & ~~$264.4 \pm 18.6$~~ & ~~~~$8.1 \pm 0.5$ & $-19.6 \pm 0.3$~~ \\
HIP\,98194	     & $70.219 \pm 0.031$   & $2457670.27 \pm 0.31$ & $0.351 \pm 0.008$ & $293.9 \pm 1.5$~~    & ~~$47.5 \pm 0.4$  & $-21.0 \pm 0.3$~~ \\
\hline \\
	\end{tabular}  }
	\label{elements}
\end{table*}
\begin{figure*}
 \includegraphics[width=0.493\textwidth]{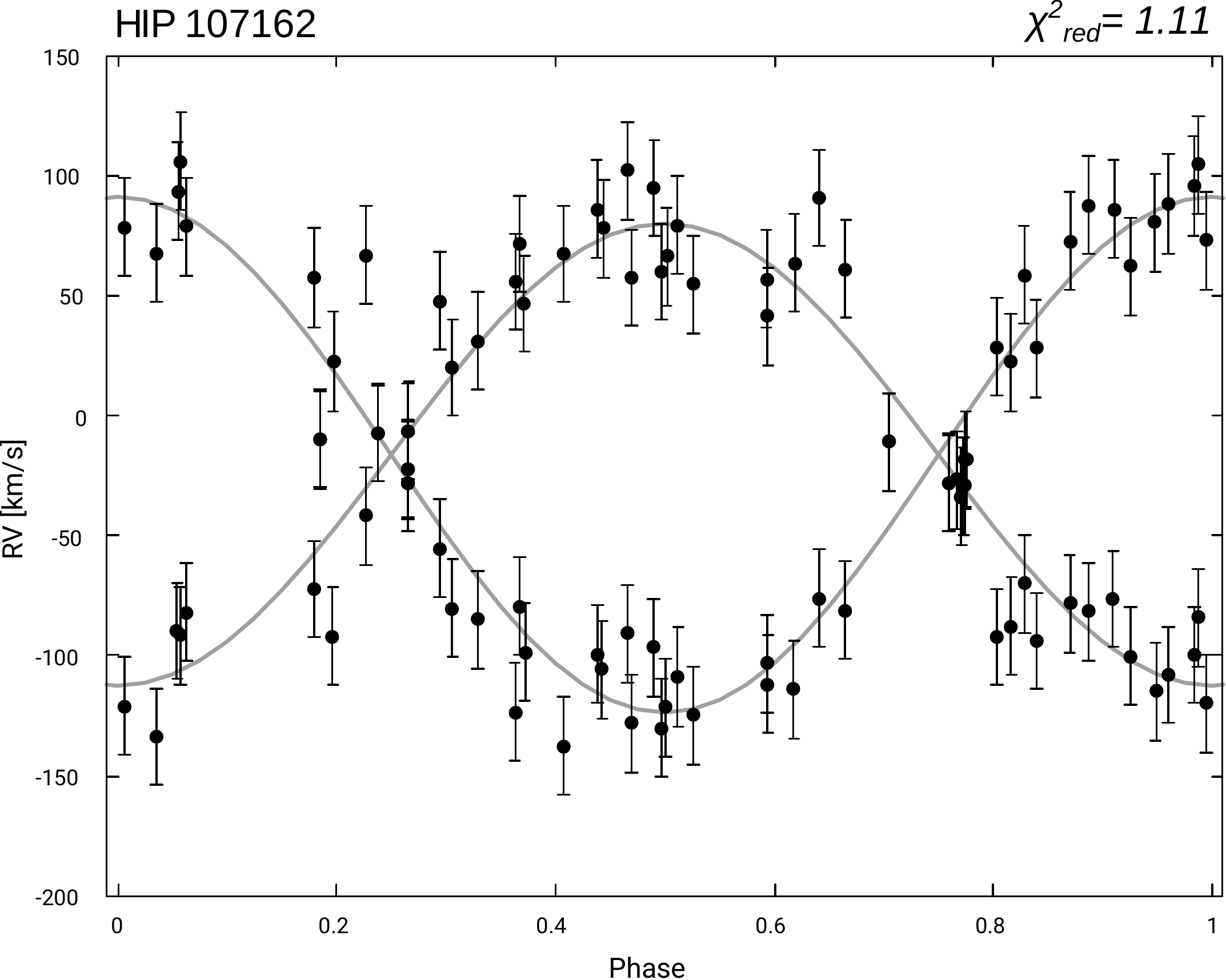}\hspace{2.5mm}\includegraphics[width=0.493\textwidth]{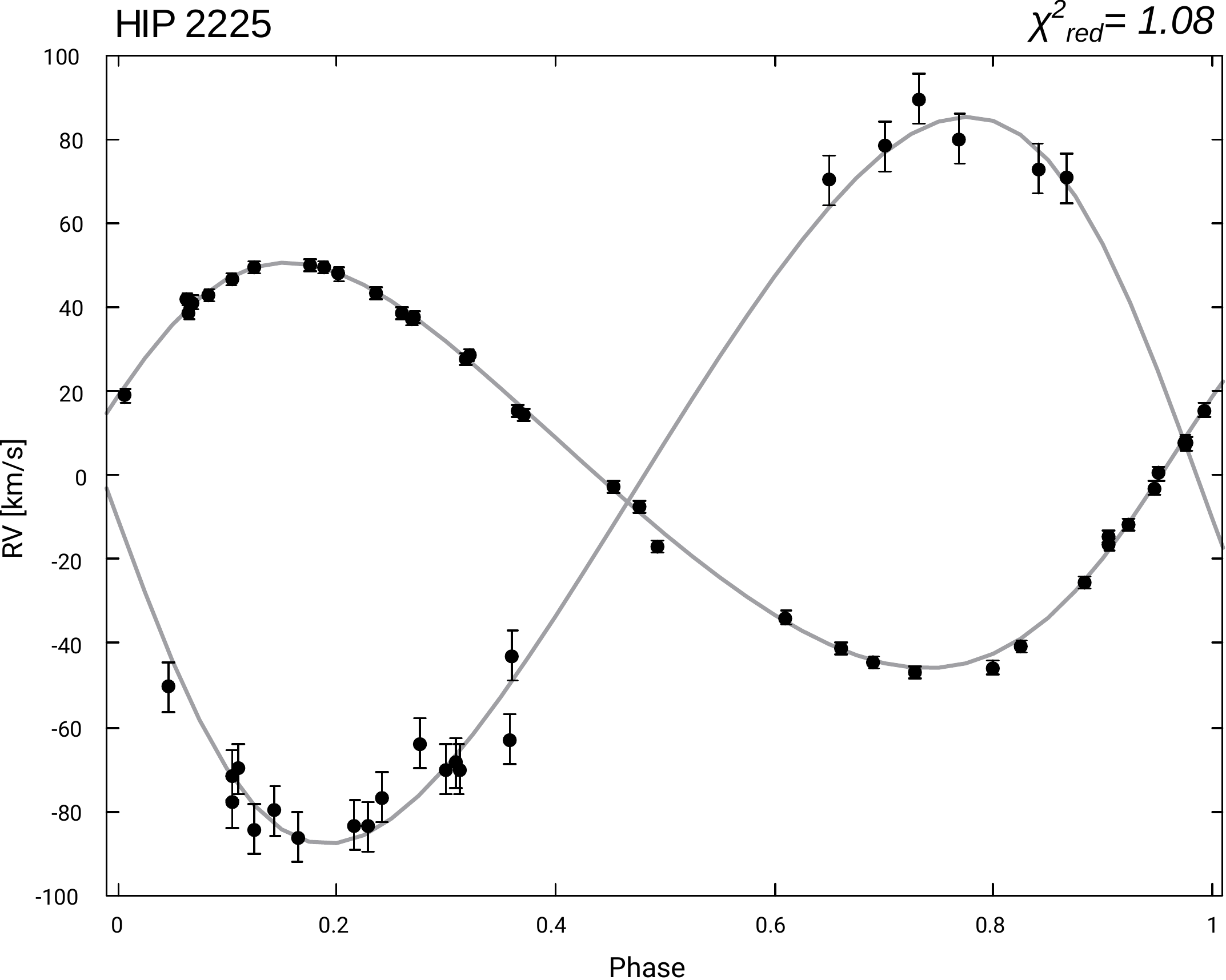}
  	\caption{The phase-folded RV curves of the double-lined spectroscopic binaries HIP\,107162, and HIP\,2225. The RV measurements are shown with their uncertainties together with the derived best fitting Keplerian orbital solutions (grey lines).}
 \label{fig:elements_sb2}
\end{figure*}

The RV variation of a component of a binary star due to Keplerian orbital motion around the barycenter of the stellar system is described by

\begin{equation}
RV = K \cdot [e \cdot \cos(\omega) + \cos(\omega + \nu)] + \gamma
\end{equation}

where $K$ is the semi-amplitude, $\nu$ the true anomaly of the component, $\omega$ the argument of periastron, $e$ the eccentricity of its orbit and $\gamma$ the systematic velocity of the stellar system.

The semi-amplitude

\begin{equation}
K = \frac{2\pi \cdot a \sin(i)}{P \cdot \sqrt{1-e^2}}
\end{equation}

depends on the minimum semi-major axis $a \sin(i)$, the orbital period $P$, and the orbital eccentricity of the component of the spectroscopic binary system.

For all spectroscopic binaries the orbital elements together with their uncertainties are determined by fitting Keplerian orbital solutions on the obtained RV measurements of the systems, using the spectroscopic binary solver (\cite{johnson}). The derived orbital solutions are illustrated as grey lines in the phase-folded RV curves in Fig.\,\ref{fig:elements}, and their orbital elements are summarized in Tab.\,\ref{elements}. The orbital solutions all exhibit reduced chi-squared values of $\chi^2_{red} \sim 1$ and their semi-amplitudes are at least three times larger than the given RV uncertainties, i.e. the FLECHAS spectroscopy significantly detects the Keplerian motion of all binary systems, monitored in this project.

For HIP\,30247 ($e = 0.004 \pm 0.003$), HIP\,23040 ($e = 0.000 \pm 0.006$) and HIP\,107162 ($e = 0.012 \pm 0.035$) the derived orbital solutions are consistent with circular orbits. Therefore, for these particular systems the orbit fitting was repeated, whereby the eccentricity was fixed to $e=0$.

In order to check the stability of the determined orbital solutions over the whole span of time, covered by our RV monitoring project, at first we derived the orbital solutions of all systems for both observing epochs separately. For the majority of all systems, these orbital solutions agree with each other within their uncertainties. Hence, for these systems we use all available RV data of both observing epochs to determine the best fitting Keplerian orbital solution. In contrast, for HIP\,2225 a significant difference in its systematic velocity is detected between the first ($\gamma = 22.6 \pm 0.4$\,km/s) and second ($\gamma = 13.1 \pm 0.4$\,km/s) observing epoch. In order to determine an orbital solution for this particular system, based on all RV data, in each observing epoch the derived systematic velocity is subtracted from the RV measurements, resulting in a fixed systematic velocity ($\gamma=0$\,km/s) in the final orbital solution of HIP\,2225.

For HIP\,107136 and HIP\,107533 no significant RV variation could be detected. Both stars exhibit a constant RV of $-5.8 \pm 3.1$\,km/s, and $-17.2 \pm 1.2$\,km/s, respectively.

\section{Discussion}

\begin{figure*}
\resizebox{\textwidth}{!}{\includegraphics{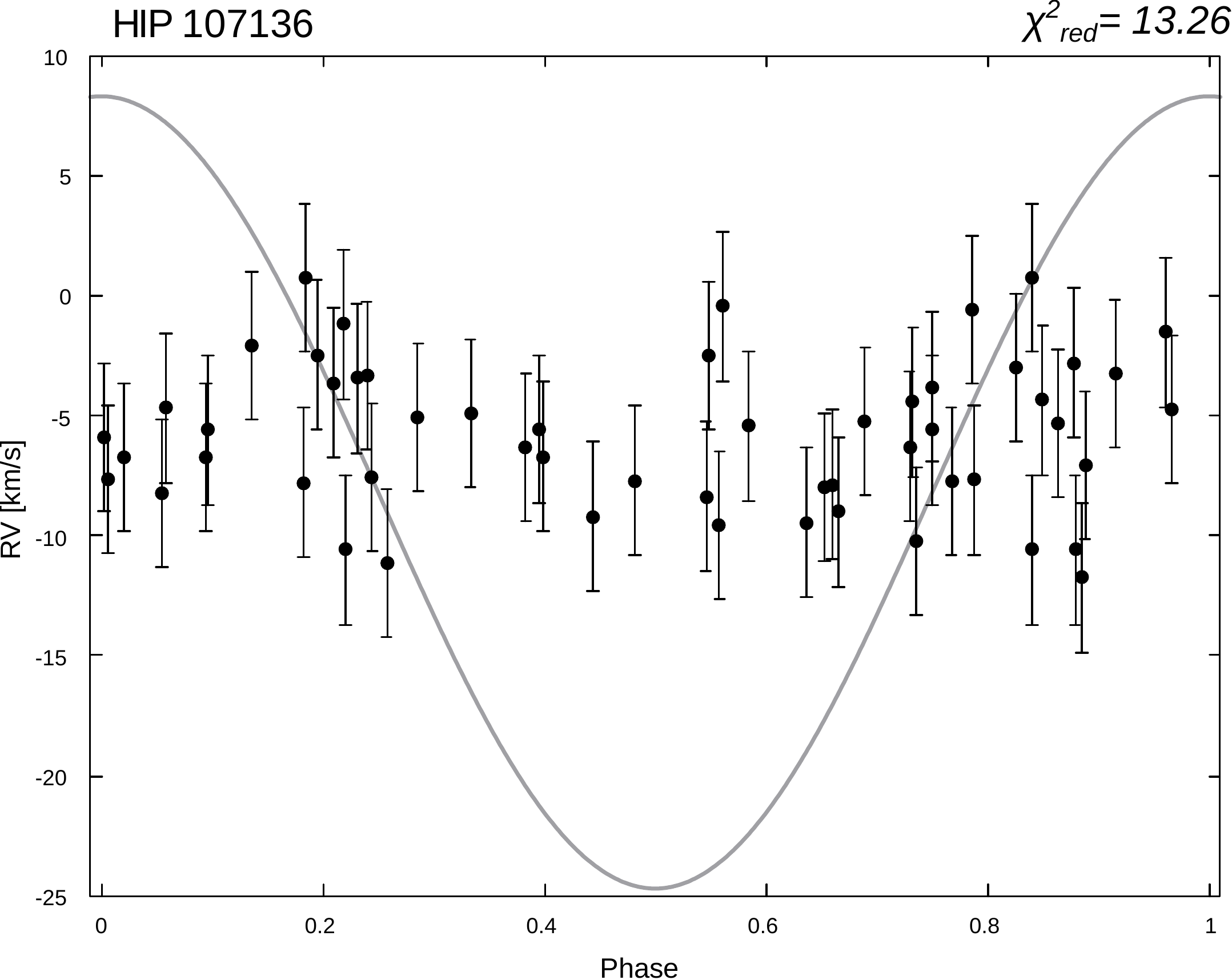}\hspace{2.5mm}\includegraphics{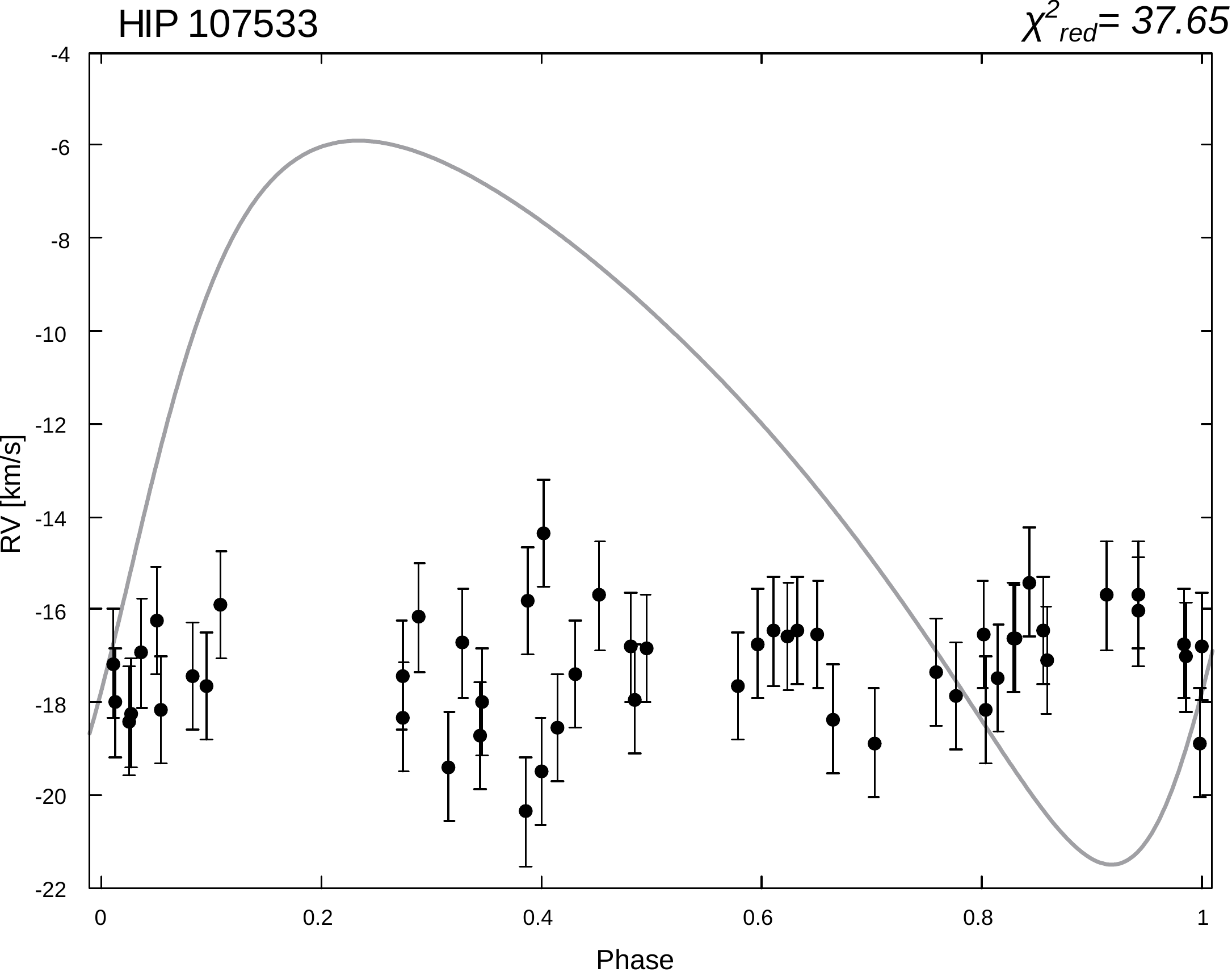}}
  	\caption{RV measurements of HIP\,107136 and HIP\,107533, taken in the course of our spectroscopic monitoring programm with FLECHAS, phase-folded with the orbital period of the stars, as given in the SB9. The orbital solutions from the SB9 are shown as grey lines, which are clearly inconsistent ($\chi^2_{\text{red}} > 13$) with the RV data of both stars, taken in this project.}
 \label{fig:elements_sb9_1}
\end{figure*}

In the course of our RV monitoring project more than 30 RV measurements could be recorded with FLECHAS for all targets, yielding homogeneously covered phase-folded RV curves of all observed spectroscopic binary systems. For the majority of these systems our RV data verify and constrain the given orbital solutions, which are listed in the SB9, and are summarized in Tab.\,\ref{table_targetproperties}.

For HIP\,25048 the relative difference between the orbital period, obtained in this project, and the one listed in the SB9 is 2.9\,\% and thus significantly larger than for the remaining 9 spectroscopic binaries, among which HIP\,77986 exhibits the largest difference of only 0.7\,\%. The eccentricity of this system deviates by 0.11 and its argument of periastron by more than $100^{\circ}$ from the SB9 value, which corresponds to the largest difference of both quantities, found among all targets, investigated in this project.

In general, the semi-amplitudes of all systems, derived here, deviate by more than 10\,\% from the amplitudes, given in the SB9. In the case of HIP\,113048 the semi-amplitude, listed in the SB9, is even twice as large as the amplitude of the system derived from our RV measurements. Only for HIP\,30247 and HIP\,23040 the obtained semi-amplitudes agree on the 3 percent level with the SB9 amplitudes.

On average, the systematic velocities of all systems, derived in this project, differ by about 2\,km/s from the velocities given in the SB9, with the largest difference of about 9\,km/s found for HIP\,107162.

HIP\,107136, and HIP\,107533 are both listed as single-lined spectroscopic binaries in the SB9 but no significant RV variation could be detected for these stars in our RV monitoring project within a span of time of 545\,days. Furthermore, as illustrated in Fig.\,\ref{fig:elements_sb9_1}, our RV measurements clearly rule out the orbital solutions of both stars, as given in the SB9, but prove that they exhibit constant RVs of $-5.8 \pm 3.1$\,km/s, and $-17.2 \pm 1.2$\,km/s, respectively.

HIP\,2225, which is classified as a single-lined spectroscopic binary in the SB9, turns out to be a actually a double-lined spectroscopic binary system, as confirmed with the RV measurements, taken in the course of our project. Furthermore, while for the majority of the observed binary stars their orbital solutions, derived for both observing epochs separately, are consistent with each other within their uncertainties, HIP\,2225 exhibits a significant variation of its systematic velocity $\Delta \gamma = -9.5 \pm 0.6$\,km/s between both observing epochs, which yields $\dot{\gamma} = -7.8 \pm 1.4\,\text{km}\,\text{s}^{-1}\,\text{yr}^{-1}$. This deviation of the systematic velocity indicates the presence of a further companion on a wider orbit in this system. For a companion on circular orbit with the radius $a$ one expects a maximal deviation of the systematic velocity

\begin{equation}
   \dot{\gamma}_{max}=MG/a^2
\end{equation}

with $M$ the total mass of the system.

Indeed, \cite{horch} detected a faint companion-candidate ($\Delta m \sim 2.4$\,mag) in the (infra)red spectral range, which is located only about $0.1$\,arcsec southwest of HIP\,2225. If this candidate is a bound companion of the star its mass can be approximated using its magnitude difference to HIP\,2225, as well as the spectral type of the star (A2V, as given in the SB9). According to the spectral type - effective temperature relation from \cite{pecaut} this yields a mass of the companion-candidate of about $1.1\,_{M_\odot}$. With the mass of HIP\,2225 ($2.26\,_{M_\odot}$, as determined by \cite{allende}) and the derived mass-ratio of the spectroscopic binary ($\sim$0.56, see Tab.\,\ref{ma_sb2}), the total mass of the HIP\,2225 system can be estimated to be $4.6\,_{M_\odot}$.

Hence, assuming $\dot{\gamma}=\dot{\gamma}_{max}$, the orbital radius of the wide companion is about 10.5\,au, which corresponds to an orbital period of $\sim$16\,yr. At the distance of HIP\,2225 ($\pi = 11.2634$\,mas, \cite{gaia}) this yields a maximal projected angular separation between the companion and the star of about 0.12\,arcsec, fully consistent with the observed angular separation of the companion-candidate, directly imaged by \cite{horch}. Therefore, we consider this candidate as a bound companion of HIP\,2225, which induces the detected deviation of the systematic velocity of this system. Furthermore, the observed higher eccentricity of the close spectroscopic binary might also indicate the presence of an additional companion of HIP\,2225. In order to further constrain the orbital parameters of this wide companion, which should be denoted as HIP\,2225\,C, additional RV monitoring and high contrast imaging observations are needed, which have to be carried out within the next one and a half decades.

Finally, for all observed single-lined spectroscopic binaries their mass-function $f(M)$ and the minimum semi-major axis $a\sin(i)$ can be derived from the determined orbital elements of these systems

\begin{equation}
  f(M) = \frac{\left(1 - e^2 \right)^{\frac{3}{2}} \cdot P \cdot K^3}{2 \pi \cdot G}
\end{equation}

\begin{equation}
  a\sin(i) = \frac{K \cdot P \cdot \sqrt{1- e^2}}{2 \pi}
\end{equation}

which are listed with their uncertainties in Tab.\,\ref{fma}.

\begin{table}[h!]
\caption{The derived mass function $f(M)$ and minimum semi-major axis $a\sin(i)$ of all single-lined spectroscopic binary systems, observed in this project.}
\centering
\begin{tabular}{ccccc}
\hline
Target       & $f(M)$                &  $a\sin(i)$\\
             & $[\text{M}_{\odot}]$  & $[$au$]$\\
\hline
HIP\,23040	 & $0.0171  \pm 0.0002$  & $0.0125 \pm 0.0001$\\
HIP\,30247	 & $0.0224  \pm 0.0002$  & $0.0192 \pm 0.0001$\\
HIP\,113048	 & $0.00021 \pm 0.00002$ & $0.0097 \pm 0.0004$\\
HIP\,25048	 & $0.1644  \pm 0.0069$  & $0.1136 \pm 0.0016$\\
HIP\,85829	 & $0.0050  \pm 0.0002$  & $0.0379 \pm 0.0005$\\
HIP\,77986	 & $0.0023  \pm 0.0005$  & $0.0331 \pm 0.0022$\\
HIP\,98194	 & $0.6392  \pm 0.0186$  & $0.2869 \pm 0.0028$\\
\hline
\end{tabular}                                              		
	\label{fma}
\end{table}

In addition, for the double-lined spectroscopic binaries the minimum-masses of their components

\begin{equation}
M_{1/2}\sin^3(i)=\frac{(1-e^2)^{3/2}P}{2\pi G}K_{2/1}(K_1+K_2)^2
\end{equation}

as well as their minimum semi-major axes can be determined, which are summarized with their uncertainties in Tab.\,\ref{ma_sb2}.

\begin{table}[h!]
\caption{Derived minimum-masses $M\sin^3(i)$ and semi-major axes $a\sin(i)$ for both components of the observed double-lined spectroscopic binary systems HIP\,107162, and HIP\,2225.}
\centering
\resizebox{1\linewidth}{!}{
\begin{tabular}{ccc}
\hline
Target          & $M\sin^3(i)$         & $a\sin(i)$\\
Component       & $[\text{M}_{\odot}]$ & $[$au$]$\\
\hline
HIP\,107162\,A  & $0.798 \pm 0.001$    & $0.0153 \pm 0.0003$ \\
HIP\,107162\,B  & $0.715 \pm 0.001$    & $0.0171 \pm 0.0003$ \\
HIP\,2225\,A    & $0.628 \pm 0.016$    & $0.0176 \pm 0.0006$ \\
HIP\,2225\,B    & $0.354 \pm 0.016$    & $0.0311 \pm 0.0006$ \\
\hline
\end{tabular}}                                              		
	\label{ma_sb2}
\end{table}

The single-lined spectroscopic binaries exhibit mass-functions in the range between 0.0002 and 0.64\,M$_{\odot}$ and their minimum semi-major axes range between about 0.01 and 0.29\,au.

While the double-lined spectroscopic binary HIP\,107162 consists of two comparably massive stars on circular orbits (consistent with the result, given in the SB9), the components of HIP\,2225 exhibit a mass-ratio of $0.56 \pm 0.03$, and revolve around their barycenter on close ($a < 0.1$\,au) and slightly eccentric ($e \sim 0.14$) orbits with a period of about 4\,days. In addition, a further companion, HIP\,2225\,C, could spectroscopically be identified in this system, which orbits the close binary with a period of about 16\,yr, or more.

The RV monitoring project of spectroscopic binaries, presented here, is part of an ongoing program, carried out at the University Observatory Jena. In the future we will continue to take RV measurements of the remaining targets of this program until precise orbital solutions for these systems can be derived from our RV data. Theses RV measurements and orbital solutions, together with those presented in this paper, will be made available online at VizieR (\cite{ochsenbein}).

\begin{acknowledgement}
We thank all observers, who have been involved in some of the observations of this project at the University Observatory Jena, in particular H. Gilbert, T. Zehe, A. Trepanovski, and F. Schiefeneder. This publication makes use of data products of the \texttt{VizieR} databases, as well as \texttt{SIMBAD}, both operated at CDS, Strasbourg, France.

\end{acknowledgement}

\newpage
\appendix\section{Radial velocity measurements}

\begin{center}
\vspace{0.8cm}
 \tablefirsthead{%
\hline
Date of Obs.         & \multicolumn{1}{c}{$\text{RV}_1$} & \multicolumn{1}{c}{$\text{RV}_2$} \\
		$\text{BJD}-2450000$ & \multicolumn{1}{c}{$[$km/s$]$} & \multicolumn{1}{c}{$[$km/s$]$} \\
		\hline}
\tablehead{%
\multicolumn{1}{l}{Continued}\\
\hline
Date of Obs.         & \multicolumn{1}{c}{$\text{RV}_1$} & \multicolumn{1}{c}{$\text{RV}_2$} \\
		$\text{BJD}-2450000$ & \multicolumn{1}{c}{$[$km/s$]$} & \multicolumn{1}{c}{$[$km/s$]$} \\
		\hline}
\tabletail{%
\hline	}
\tablelasttail{\hline}
\tablecaption{RV measurements of HIP\,107162}
\begin{supertabular}{c @{\hspace{0.9cm}} r r}
$7728.27661$ & $-26.8 \pm 20.2$ & $-26.8 \pm 20.4$ \\
$7729.20791$ & $+20.2 \pm 20.2$ & $-80.1 \pm 20.4$ \\
$7743.26094$ & $+86.1 \pm 20.2$ & $-99.3 \pm 20.4$ \\
$7744.20421$ & $-99.6 \pm 20.2$ & $+95.7 \pm 20.4$ \\
$7760.24442$ & $-22.4 \pm 20.2$ & $-22.4 \pm 20.4$ \\
$7764.27020$ & $+41.5 \pm 20.2$ & $-111.8 \pm 20.4$ \\
$7773.22328$ & $-18.3 \pm 20.2$ & $-18.3 \pm 20.4$ \\
$7775.25176$ & $-114.8 \pm 20.2$& $+80.5 \pm 20.4$ \\
$7775.33198$ & $-119.9 \pm 20.2$& $+73.1 \pm 20.4$ \\
$7776.22491$ & $+79.2 \pm 20.2$ & $-108.8 \pm 20.4$ \\
$7780.22873$ & $-69.9 \pm 20.2$ & $+58.6 \pm 20.4$ \\
$7782.23129$ & $-84.1 \pm 20.2$ & $+104.8 \pm 20.4$ \\
$7782.71285$ & $-06.4 \pm 20.2$ & $-06.4 \pm 20.4$ \\
$7798.26671$ & $-27.9 \pm 20.2$ & $-27.9 \pm 20.4$ \\
$7800.24044$ & $+67.5 \pm 20.2$ & $-137.7 \pm 20.4$ \\
$7840.59677$ & $-28.0 \pm 20.2$ & $-28.0 \pm 20.4$ \\
$7843.60805$ & $+66.4 \pm 20.2$ & $-121.6 \pm 20.4$ \\
$7874.49172$ & $+46.8 \pm 20.2$ & $-98.4 \pm 20.4$ \\
$7880.48457$ & $-93.9 \pm 20.2$ & $+28.2 \pm 20.4$ \\
$8151.17594$ & $+57.3 \pm 20.2$ & $-128.0 \pm 20.4$ \\
$8151.22274$ & $+59.9 \pm 20.2$ & $-130.0 \pm 20.4$ \\
$8151.27354$ & $+54.8 \pm 20.2$ & $-124.9 \pm 20.4$ \\
$8155.23215$ & $-87.8 \pm 20.2$ & $+22.2 \pm 20.4$ \\
$8172.63542$ & $-81.7 \pm 20.2$ & $+87.7 \pm 20.4$ \\
$8173.63656$ & $+102.1 \pm 20.2$ & $-90.7 \pm 20.4$ \\
$8174.65402$ & $-89.8 \pm 20.2$ & $+93.4 \pm 20.4$ \\
$8178.65121$ & $+71.5 \pm 20.2$ & $-79.6 \pm 20.4$ \\
$8197.59560$ & $+31.2 \pm 20.2$ & $-84.8 \pm 20.4$ \\
$8202.53067$ & $-09.9 \pm 20.2$ & $-09.9 \pm 20.4$ \\
$8207.50535$ & $-81.9 \pm 20.2$ & $+78.7 \pm 20.4$ \\
$8215.52609$ & $-11.0 \pm 20.2$ & $-11.0 \pm 20.4$ \\
$8216.54567$ & $+47.9 \pm 20.2$ & $-55.2 \pm 20.4$ \\
$8217.54034$ & $-78.3 \pm 20.2$ & $+72.6 \pm 20.4$ \\
$8218.53161$ & $+78.0 \pm 20.2$ & $-105.6 \pm 20.4$ \\
$8219.50427$ & $-120.9 \pm 20.2$ & $+78.4 \pm 20.4$ \\
$8220.51770$ & $+56.9 \pm 20.2$ & $-103.0 \pm 20.4$ \\
$8228.52748$ & $-41.8 \pm 20.2$ & $+66.9 \pm 20.4$ \\
$8229.52273$ & $-92.0 \pm 20.2$ & $+28.8 \pm 20.4$ \\
$8230.49063$ & $+55.6 \pm 20.2$ & $-123.4 \pm 20.4$ \\
$8238.43508$ & $-108.1 \pm 20.2$ & $+88.2 \pm 20.4$ \\
$8240.54190$ & $-72.1 \pm 20.2$ & $+57.4 \pm 20.4$ \\
$8243.53264$ & $-76.4 \pm 20.2$ & $+86.1 \pm 20.4$ \\
$8244.53565$ & $+94.9 \pm 20.2$ & $-96.5 \pm 20.4$ \\
$8245.51672$ & $-91.6 \pm 20.2$ & $+106.0 \pm 20.4$ \\
$8246.52495$ & $+90.8 \pm 20.2$ & $-76.0 \pm 20.4$ \\
$8259.58250$ & $-91.8 \pm 20.2$ & $+22.6 \pm 20.4$ \\
$8260.57475$ & $-33.6 \pm 20.2$ & $-33.6 \pm 20.4$ \\
$8264.48912$ & $-133.6 \pm 20.2$ & $+67.6 \pm 20.4$ \\
$8265.49587$ & $+63.7 \pm 20.2$ & $-114.0 \pm 20.4$ \\
$8267.49156$ & $-29.4 \pm 20.2$ & $-29.4 \pm 20.4$ \\
$8269.48234$ & $-100.3 \pm 20.2$ & $+62.3 \pm 20.4$ \\
$8272.48952$ & $+61.0 \pm 20.2$ & $-81.2 \pm 20.4$ \\
$8273.47945$ & $-07.3 \pm 20.2$ & $-07.3 \pm 20.4$ \\
\end{supertabular}
\label{tab:HIP107162}
\end{center}

\begin{center}
\vspace{0.8cm}
 \tablefirsthead{%
\hline
Date of Obs.         & \multicolumn{1}{c}{RV} \\
		$\text{BJD}-2450000$ & \multicolumn{1}{c}{$[$km/s$]$} \\
		\hline}
\tablehead{%
\multicolumn{1}{l}{Continued}\\
\hline
Date of Obs.         & \multicolumn{1}{c}{RV} \\
		$\text{BJD}-2450000$ & \multicolumn{1}{c}{$[$km/s$]$} \\
		\hline}
\tabletail{%
\hline	}
\tablelasttail{\hline}
\tablecaption{RV measurements of HIP\,23040}
\begin{supertabular}{c @{\hspace{2cm}} r}
$7729.47686$ & $+16.3 \pm 1.1$ \\
$7743.20867$ & $-40.7 \pm 1.1$ \\
$7743.64393$ & $-44.2 \pm 1.1$ \\
$7744.24491$ & $-24.3 \pm 1.1$ \\
$7744.52109$ & $-9.5 \pm 1.1$ \\
$7759.52539$ & $-35.6 \pm 1.1$ \\
$7760.27312$ & $+2.7 \pm 1.1$ \\
$7764.21819$ & $+5.2 \pm 1.1$ \\
$7773.29456$ & $+8.5 \pm 1.1$ \\
$7775.23967$ & $-27.3 \pm 1.1$ \\
$7775.31930$ & $-24.7 \pm 1.1$ \\
$7776.27412$ & $+20.6 \pm 1.1$ \\
$7776.56061$ & $+23.0 \pm 1.1$ \\
$7780.25179$ & $+22.0 \pm 1.1$ \\
$7780.61282$ & $+23.4 \pm 1.1$ \\
$7782.27518$ & $-43.8 \pm 1.1$ \\
$7782.36730$ & $-45.5 \pm 1.1$ \\
$7786.39259$ & $-45.0 \pm 1.1$ \\
$7798.35257$ & $-36.7 \pm 1.1$ \\
$7798.50655$ & $-31.2 \pm 1.1$ \\
$7799.35985$ & $+14.2 \pm 1.1$ \\
$7799.53053$ & $+19.6 \pm 1.1$ \\
$7800.35056$ & $+14.8 \pm 1.1$ \\
$7800.53099$ & $+5.9 \pm 1.1$ \\
$7812.30577$ & $-1.0 \pm 1.1$ \\
$7826.41535$ & $+8.1 \pm 1.1$ \\
$7826.45006$ & $+9.6 \pm 1.1$ \\
$7840.30417$ & $-44.8 \pm 1.1$ \\
$7840.49964$ & $-43.8 \pm 1.1$ \\
$7841.30362$ & $-25.7 \pm 1.1$ \\
$7841.40664$ & $-22.3 \pm 1.1$ \\
$7843.30108$ & $+4.8 \pm 1.1$ \\
$7843.46655$ & $-4.6 \pm 1.1$ \\
$7864.33996$ & $-38.0 \pm 1.1$ \\
$7874.37330$ & $+5.4 \pm 1.1$ \\
$7876.40554$ & $-17.4 \pm 1.1$ \\
$8146.28347$ & $+5.0 \pm 1.1$ \\
$8149.35755$ & $+24.6 \pm 1.1$ \\
$8151.38000$ & $-44.0 \pm 1.1$ \\
$8151.59191$ & $-43.1 \pm 1.1$ \\
$8154.45375$ & $-18.2 \pm 1.1$ \\
$8154.51927$ & $-19.7 \pm 1.1$ \\
$8154.56530$ & $-24.0 \pm 1.1$ \\
$8158.50489$ & $-25.9 \pm 1.1$ \\
$8161.43027$ & $+21.1 \pm 1.1$ \\
$8162.37449$ & $-24.7 \pm 1.1$ \\
$8163.53019$ & $-34.0 \pm 1.1$ \\
$8164.50423$ & $+14.0 \pm 1.1$ \\
$8166.57512$ & $-39.2 \pm 1.1$ \\
$8168.32079$ & $+12.9 \pm 1.1$ \\
$8168.44797$ & $+18.0 \pm 1.1$ \\
$8170.42659$ & $-37.0 \pm 1.1$ \\
$8170.56073$ & $-39.8 \pm 1.1$ \\
$8171.38046$ & $-31.7 \pm 1.1$ \\
$8171.54108$ & $-24.7 \pm 1.1$ \\
$8172.48056$ & $+21.5 \pm 1.1$ \\
$8173.33911$ & $+12.1 \pm 1.1$ \\
$8174.33208$ & $-38.2 \pm 1.1$ \\
$8174.49104$ & $-42.1 \pm 1.1$ \\
$8175.35661$ & $-27.1 \pm 1.1$ \\
$8175.52627$ & $-18.9 \pm 1.1$ \\
$8176.40893$ & $+21.5 \pm 1.1$ \\
$8177.31105$ & $+7.5 \pm 1.1$ \\
$8177.42488$ & $+2.3 \pm 1.1$ \\
$8178.34185$ & $-42.4 \pm 1.1$ \\
$8178.42082$ & $-43.2 \pm 1.1$ \\
$8179.38634$ & $-19.0 \pm 1.1$ \\
$8179.56806$ & $-9.9 \pm 1.1$ \\
$8183.35664$ & $-14.5 \pm 1.1$ \\
$8190.42398$ & $-42.6 \pm 1.1$ \\
$8197.29127$ & $-22.6 \pm 1.1$ \\
$8202.30717$ & $-37.3 \pm 1.1$ \\
$8202.46439$ & $-32.1 \pm 1.1$ \\
$8213.42304$ & $-43.2 \pm 1.1$ \\
$8214.44174$ & $-14.8 \pm 1.1$ \\
$8215.28578$ & $+23.7 \pm 1.1$ \\
$8216.39092$ & $-5.8 \pm 1.1$ \\
$8217.28049$ & $-43.4 \pm 1.1$ \\
$8218.33471$ & $-14.1 \pm 1.1$ \\
$8218.40841$ & $-10.1 \pm 1.1$ \\
$8219.29530$ & $+22.4 \pm 1.1$ \\
$8219.42595$ & $+23.9 \pm 1.1$ \\
$8220.43200$ & $-14.1 \pm 1.1$ \\
$8223.40518$ & $+25.7 \pm 1.1$ \\
$8226.31235$ & $-2.3 \pm 1.1$ \\
$8227.31850$ & $+22.7 \pm 1.1$ \\
$8228.31137$ & $-19.3 \pm 1.1$ \\
$8229.35252$ & $-40.4 \pm 1.1$ \\
$8230.32692$ & $+6.5 \pm 1.1$ \\
$8230.42533$ & $+10.9 \pm 1.1$ \\
$8232.41707$ & $-31.1 \pm 1.1$ \\
$8236.33528$ & $-34.4 \pm 1.1$ \\
$8238.33184$ & $+16.1 \pm 1.1$ \\
$8239.35115$ & $+13.3 \pm 1.1$ \\
$8240.34454$ & $-39.6 \pm 1.1$ \\
$8243.31857$ & $+10.9 \pm 1.1$ \\
$8244.32752$ & $-40.0 \pm 1.1$ \\
$8245.33033$ & $-24.1 \pm 1.1$ \\
$8246.33149$ & $+22.3 \pm 1.1$ \\
$8247.31681$ & $+0.9 \pm 1.1$ \\
$8248.36928$ & $-45.6 \pm 1.1$ \\
$8252.33853$ & $-43.7 \pm 1.1$ \\
$8253.33877$ & $-10.2 \pm 1.1$ \\
$8264.34598$ & $-41.0 \pm 1.1$ \\
\end{supertabular}
\end{center}

\begin{center}
\vspace{0.8cm}
 \tablefirsthead{%
 \hline
Date of Obs.         & \multicolumn{1}{c}{$\text{RV}_1$} & \multicolumn{1}{c}{$\text{RV}_2$} \\
		$\text{BJD}-2450000$ & \multicolumn{1}{c}{$[$km/s$]$} & \multicolumn{1}{c}{$[$km/s$]$} \\
		\hline}
\tablehead{%
\multicolumn{1}{l}{Continued}\\
\hline
Date of Obs.         & \multicolumn{1}{c}{$\text{RV}_1$} & \multicolumn{1}{c}{$\text{RV}_2$} \\
		$\text{BJD}-2450000$ & \multicolumn{1}{c}{$[$km/s$]$} & \multicolumn{1}{c}{$[$km/s$]$} \\
		\hline}
\tabletail{%
\hline	}
\tablelasttail{\hline}
\tablecaption{RV measurements of HIP\,2225}
\begin{supertabular}{c @{\hspace{0.9cm}} r r}
$7729.29877$ & $+5.6 \pm 2.3$ & \\
$7743.19728$ & $+41.5 \pm 2.3$ & $-27.8 \pm 5.9$ \\
$7743.42810$ & $+61.1 \pm 2.3$ & $-48.9 \pm 5.9$ \\
$7744.23432$ & $+59.8 \pm 2.3$ & $-45.9 \pm 5.9$ \\
$7760.26168$ & $+51.1 \pm 2.3$ & $-20.3 \pm 5.9$ \\
$7764.20643$ & $+50.1 \pm 2.3$ & $-40.3 \pm 5.9$ \\
$7773.27365$ & $-11.3 \pm 2.3$ & $+92.9 \pm 5.9$ \\
$7775.22609$ & $+69.3 \pm 2.3$ & $-57.3 \pm 5.9$ \\
$7775.30591$ & $+72.2 \pm 2.3$ & $-63.7 \pm 5.9$ \\
$7776.26072$ & $+37.9 \pm 2.3$ & \\
$7780.23921$ & $+37.0 \pm 2.3$ & \\
$7782.26301$ & $-2.9 \pm 2.3$ & \\
$7782.35320$ & $+6.1 \pm 2.3$ & \\
$7786.38003$ & $+10.8 \pm 2.3$ & \\
$7798.33964$ & $+19.5 \pm 2.3$ & \\
$7799.34737$ & $+70.4 \pm 2.3$ & $-54.2 \pm 5.9$\\
$7800.33869$ & $+19.8 \pm 2.3$ &  	\\
$8146.26468$ & $-1.6 \pm 2.3$ &  	\\
$8149.25184$ & $-28.1 \pm 2.3$ & $+91.4 \pm 5.9$\\
$8149.37150$ & $-31.4 \pm 2.3$ & $+102.5 \pm 5.9$\\
$8151.33825$ & $+62.5 \pm 2.3$ & $-70.6 \pm 5.9$\\
$8155.24540$ & $+63.1 \pm 2.3$ & $-70.2 \pm 5.9$\\
$8162.36250$ & $+21.0 \pm 2.3$ & 	\\
$8166.22019$ & $+13.5 \pm 2.3$ & 	\\
$8168.30229$ & $+5.5 \pm 2.3$ & 	\\
$8173.25304$ & $-33.8 \pm 2.3$ & $+93.0\pm 5.9$\\
$8174.23479$ & $+20.6 \pm 2.3$ & 	\\
$8175.26104$ & $+56.5 \pm 2.3$ & $-50.7\pm 5.9$\\
$8178.25545$ & $+28.6 \pm 2.3$ & 	\\
$8179.30977$ & $+51.7 \pm 2.3$ & $-57.1 \pm 5.9$\\
$8183.31206$ & $+50.8 \pm 2.3$ & $-57.0 \pm 5.9$\\
$8197.26890$ & $-32.7 \pm 2.3$ & $+86.1 \pm 5.9$\\
$8202.28991$ & $+54.3 \pm 2.3$ & $-56.7 \pm 5.9$\\
$8269.51271$ & $+54.9 \pm 2.3$ & $-64.7 \pm 5.9$\\
$8272.52835$ & $-27.7 \pm 2.3$ & $+83.9 \pm 5.9$\\
$8273.54855$ & $+56.0 \pm 2.3$ & $-71.2 \pm 5.9$\\
\end{supertabular}
\end{center}

\begin{center}
\vspace{0.8cm}
 \tablefirsthead{%
\hline
Date of Obs.         & \multicolumn{1}{c}{RV} \\
		$\text{BJD}-2450000$ & \multicolumn{1}{c}{$[$km/s$]$} \\
		\hline}
\tablehead{%
\multicolumn{1}{l}{Continued}\\
\hline
Date of Obs.         & \multicolumn{1}{c}{RV} \\
		$\text{BJD}-2450000$ & \multicolumn{1}{c}{$[$km/s$]$} \\
		\hline}
\tabletail{%
\hline	}
\tablelasttail{\hline}
\tablecaption{RV measurements of HIP\,30247}
\begin{supertabular}{c @{\hspace{2cm}} r}
 $7729.48413$ & $-26.4 \pm 0.8$ \\
$7743.21598$ & $-29.5 \pm 0.8$ \\
$7743.65114$ & $-22.8 \pm 0.8$ \\
$7744.25195$ & $-8.4 \pm 0.8$ \\
$7744.52804$ & $-0.0 \pm 0.8$ \\
$7759.59163$ & $+32.3 \pm 0.8$ \\
$7760.29840$ & $+17.6 \pm 0.8$ \\
$7764.22588$ & $+7.5 \pm 0.8$ \\
$7773.33327$ & $+16.4 \pm 0.8$ \\
$7775.24820$ & $-29.9 \pm 0.8$ \\
$7775.32734$ & $-30.1 \pm 0.8$ \\
$7776.28259$ & $-20.3 \pm 0.8$ \\
$7776.56980$ & $-12.8 \pm 0.8$ \\
$7780.26028$ & $+4.0 \pm 0.8$ \\
$7780.60512$ & $-7.0 \pm 0.8$ \\
$7786.40730$ & $+14.7 \pm 0.8$ \\
$7798.36956$ & $+34.0 \pm 0.8$ \\
$7798.51495$ & $+34.2 \pm 0.8$ \\
$7799.37443$ & $+16.1 \pm 0.8$ \\
$7799.53873$ & $+11.6 \pm 0.8$ \\
$7800.36601$ & $-13.8 \pm 0.8$ \\
$7800.53927$ & $-19.0 \pm 0.8$ \\
$7812.32202$ & $+18.1 \pm 0.8$ \\
$7812.33992$ & $+17.6 \pm 0.8$ \\
$7826.38538$ & $-12.7 \pm 0.8$ \\
$7826.45871$ & $-15.3 \pm 0.8$ \\
$7840.29120$ & $-28.6 \pm 0.8$ \\
$7840.50940$ & $-30.3 \pm 0.8$ \\
$7841.28789$ & $-20.0 \pm 0.8$ \\
$7841.41769$ & $-16.1 \pm 0.8$ \\
$7843.35911$ & $+32.5 \pm 0.8$ \\
$7843.47680$ & $+33.1 \pm 0.8$ \\
$7864.35739$ & $+15.9 \pm 0.8$ \\
$7874.38269$ & $-6.5 \pm 0.8$ \\
$7874.47676$ & $-3.5 \pm 0.8$ \\
$7876.41930$ & $+34.2 \pm 0.8$ \\
$7880.38713$ & $-18.9 \pm 0.8$ \\
$8146.32840$ & $-28.5 \pm 0.8$ \\
$8149.33891$ & $+34.7 \pm 0.8$ \\
$8149.43473$ & $+33.2 \pm 0.8$ \\
$8151.43858$ & $-14.6 \pm 0.8$ \\
$8151.60186$ & $-19.2 \pm 0.8$ \\
$8154.46383$ & $+12.0 \pm 0.8$ \\
$8154.52869$ & $+13.6 \pm 0.8$ \\
$8154.57471$ & $+15.4 \pm 0.8$ \\
$8155.39213$ & $+32.3 \pm 0.8$ \\
$8158.52164$ & $-26.8 \pm 0.8$ \\
$8161.44858$ & $+25.1 \pm 0.8$ \\
$8162.43935$ & $+34.0 \pm 0.8$ \\
$8163.55380$ & $+11.7 \pm 0.8$ \\
$8164.54241$ & $-16.7 \pm 0.8$ \\
$8166.56272$ & $-13.9 \pm 0.8$ \\
$8168.35845$ & $+31.9 \pm 0.8$ \\
$8168.45663$ & $+31.7 \pm 0.8$ \\
$8170.45984$ & $-0.4 \pm 0.8$ \\
$8170.57111$ & $-3.8 \pm 0.8$  \\
$8171.41264$ & $-24.2 \pm 0.8$ \\
$8171.55471$ & $-27.0 \pm 0.8$ \\
$8172.49136$ & $-27.9 \pm 0.8$ \\
$8173.51753$ & $-2.0 \pm 0.8$ \\
$8173.58668$ & $-1.0 \pm 0.8$ \\
$8174.34104$ & $+22.7 \pm 0.8$ \\
$8174.50037$ & $+26.1 \pm 0.8$ \\
$8175.36652$ & $+34.0 \pm 0.8$ \\
$8176.41855$ & $+15.7 \pm 0.8$ \\
$8176.60954$ & $+10.4 \pm 0.8$ \\
$8177.32806$ & $-11.2 \pm 0.8$ \\
$8177.43395$ & $-13.0 \pm 0.8$ \\
$8178.36580$ & $-29.2 \pm 0.8$ \\
$8178.46354$ & $-30.2 \pm 0.8$ \\
$8179.50388$ & $-17.1 \pm 0.8$ \\
$8179.57736$ & $-15.8 \pm 0.8$ \\
$8183.36692$ & $+2.3 \pm 0.8$ \\
$8190.43306$ & $-14.3 \pm 0.8$ \\
$8197.30045$ & $-23.0 \pm 0.8$ \\
$8202.31573$ & $+20.3 \pm 0.8$ \\
$8202.47888$ & $+15.6 \pm 0.8$ \\
$8213.44009$ & $+24.1 \pm 0.8$ \\
$8214.46228$ & $+33.3 \pm 0.8$ \\
$8215.34231$ & $+18.9 \pm 0.8$ \\
$8216.41457$ & $-13.4 \pm 0.8$ \\
$8217.28883$ & $-29.6 \pm 0.8$ \\
$8218.34449$ & $-22.0 \pm 0.8$ \\
$8218.41760$ & $-18.7 \pm 0.8$ \\
$8219.37338$ & $+7.7 \pm 0.8$ \\
$8219.43695$ & $+9.9 \pm 0.8$ \\
$8223.41433$ & $-24.1 \pm 0.8$ \\
$8226.32107$ & $+21.8 \pm 0.8$ \\
$8227.37119$ & $+34.9 \pm 0.8$ \\
$8228.32089$ & $+20.6 \pm 0.8$ \\
$8229.36233$ & $-12.8 \pm 0.8$ \\
$8230.33596$ & $-30.0 \pm 0.8$ \\
$8230.41585$ & $-29.9 \pm 0.8$ \\
$8232.40959$ & $+9.1 \pm 0.8$ \\
$8236.34385$ & $-25.0 \pm 0.8$ \\
$8238.35157$ & $-7.3 \pm 0.8$ \\
$8239.37004$ & $+23.3 \pm 0.8$ \\
$8240.36838$ & $+34.1 \pm 0.8$ \\
$8243.33674$ & $-29.5 \pm 0.8$ \\
$8244.34366$ & $-20.4 \pm 0.8$ \\
$8245.34581$ & $+8.1 \pm 0.8$ \\
$8246.34864$ & $+31.1 \pm 0.8$ \\
$8247.32548$ & $+28.4 \pm 0.8$ \\
$8248.37805$ & $+3.4 \pm 0.8$ \\
$8252.34703$ & $+20.1 \pm 0.8$ \\
$8253.34657$ & $+34.0 \pm 0.8$ \\
$8259.33691$ & $+30.6 \pm 0.8$ \\
$8260.33902$ & $+28.5 \pm 0.8$ \\
$8264.35463$ & $-6.5 \pm 0.8$ \\
$8265.35050$ & $+21.7 \pm 0.8$ \\
$8267.35086$ & $+16.0 \pm 0.8$ \\
\end{supertabular}
 \label{hip30247}
\end{center}

\begin{center}
\vspace{0.8cm}
  \tablefirsthead{%
\hline
Date of Obs.         & \multicolumn{1}{c}{RV} \\
		$\text{BJD}-2450000$ & \multicolumn{1}{c}{$[$km/s$]$} \\
		\hline}
\tablehead{%
\multicolumn{1}{l}{Continued}\\
\hline
Date of Obs.         & \multicolumn{1}{c}{RV} \\
		$\text{BJD}-2450000$ & \multicolumn{1}{c}{$[$km/s$]$} \\
		\hline}
\tabletail{%
\hline	}
\tablelasttail{\hline}
\tablecaption{RV measurements of HIP\,113048}
\begin{supertabular}{c @{\hspace{2cm}} r}
$7728.36223$ & $+2.6 \pm 0.6$ \\
$7729.30555$ & $+2.1 \pm 0.6$ \\
$7743.27670$ & $+10.6 \pm 0.6$ \\
$7744.22614$ & $+10.9 \pm 0.6$ \\
$7760.25264$ & $+7.2 \pm 0.6$ \\
$7764.28884$ & $+9.3 \pm 0.6$ \\
$7773.26485$ & $+6.4 \pm 0.6$ \\
$7775.28751$ & $+3.2 \pm 0.6$ \\
$7776.25211$ & $+2.1 \pm 0.6$ \\
$7798.29252$ & $+4.5 \pm 0.6$ \\
$7800.29643$ & $+3.1 \pm 0.6$ \\
$7840.60435$ & $+10.6 \pm 0.6$ \\
$7843.62216$ & $+12.0 \pm 0.6$ \\
$7874.50114$ & $+2.9 \pm 0.6$ \\
$8151.24325$ & $+9.4 \pm 0.6$ \\
$8155.31430$ & $+11.8 \pm 0.6$ \\
$8174.24593$ & $+7.7 \pm 0.6$ \\
$8175.24795$ & $+7.6 \pm 0.6$ \\
$8197.66505$ & $+7.7 \pm 0.6$ \\
$8202.63997$ & $+11.7 \pm 0.6$ \\
$8207.61953$ & $+8.6 \pm 0.6$ \\
$8215.62892$ & $+2.9 \pm 0.6$ \\
$8216.64021$ & $+4.9 \pm 0.6$ \\
$8217.63502$ & $+6.0 \pm 0.6$ \\
$8218.62702$ & $+5.7 \pm 0.6$ \\
$8219.60784$ & $+6.9 \pm 0.6$ \\
$8220.62113$ & $+7.7 \pm 0.6$ \\
$8228.63059$ & $+11.6 \pm 0.6$ \\
$8229.61740$ & $+10.3 \pm 0.6$ \\
$8230.59348$ & $+10.6 \pm 0.6$ \\
$8238.59246$ & $+1.9 \pm 0.6$ \\
$8240.47772$ & $+3.6 \pm 0.6$ \\
$8243.49387$ & $+5.9 \pm 0.6$ \\
$8244.48957$ & $+6.9 \pm 0.6$ \\
$8245.45214$ & $+7.0 \pm 0.6$ \\
$8246.50147$ & $+7.6 \pm 0.6$ \\
$8264.42488$ & $+3.2 \pm 0.6$ \\
$8265.47786$ & $+3.8 \pm 0.6$ \\
$8267.47378$ & $+5.6 \pm 0.6$ \\
$8269.41918$ & $+6.4 \pm 0.6$ \\
$8272.43316$ & $+9.3 \pm 0.6$ \\
$8273.41615$ & $+9.5 \pm 0.6$ \\
\end{supertabular}
\end{center}

\begin{center}
\vspace{0.8cm}
 \tablefirsthead{%
\hline
Date of Obs.         & \multicolumn{1}{c}{RV} \\
		$\text{BJD}-2450000$ & \multicolumn{1}{c}{$[$km/s$]$} \\
		\hline}
\tablehead{%
\multicolumn{1}{l}{Continued}\\
\hline
Date of Obs.         & \multicolumn{1}{c}{RV} \\
		$\text{BJD}-2450000$ & \multicolumn{1}{c}{$[$km/s$]$} \\
		\hline}
\tabletail{%
\hline	}
\tablelasttail{\hline}
\tablecaption{RV measurements of HIP\,25048}
\begin{supertabular}{c @{\hspace{2cm}} r}
$7728.39985$ & $+38.9 \pm 2.8$ \\
$7729.50042$ & $+29.0 \pm 2.8$ \\
$7744.23421$ & $-17.2 \pm 2.8$ \\
$7744.50516$ & $-8.8 \pm 2.8$ \\
$7759.58410$ & $+53.2 \pm 2.8$ \\
$7760.28077$ & $+53.2 \pm 2.8$ \\
$7764.24484$ & $+35.3 \pm 2.8$ \\
$7773.32607$ & $-16.6 \pm 2.8$ \\
$7775.26858$ & $-22.3 \pm 2.8$ \\
$7776.59037$ & $-15.0 \pm 2.8$ \\
$7780.26722$ & $-3.0 \pm 2.8$ \\
$7782.28313$ & $+5.3 \pm 2.8$ \\
$7786.40001$ & $+25.8 \pm 2.8$ \\
$7798.36115$ & $+32.3 \pm 2.8$ \\
$7799.36708$ & $+22.8 \pm 2.8$ \\
$7800.35821$ & $+20.1 \pm 2.8$ \\
$7812.31385$ & $-12.1 \pm 2.8$ \\
$7826.42281$ & $+50.9 \pm 2.8$ \\
$7840.29714$ & $-8.0 \pm 2.8$ \\
$7841.29693$ & $-17.5 \pm 2.8$ \\
$7843.35039$ & $-13.6 \pm 2.8$ \\
$7849.42947$ & $-2.7 \pm 2.8$ \\
$7864.34770$ & $+50.3 \pm 2.8$ \\
$8146.30295$ & $+12.5 \pm 2.8$ \\
$8149.27496$ & $-6.6 \pm 2.8$ \\
$8151.43120$ & $-12.6 \pm 2.8$ \\
$8154.47172$ & $-16.0 \pm 2.8$ \\
$8158.51362$ & $-10.5 \pm 2.8$ \\
$8161.44113$ & $-0.8 \pm 2.8$ \\
$8162.41505$ & $+12.7 \pm 2.8$ \\
$8163.54639$ & $+14.7 \pm 2.8$ \\
$8164.53386$ & $+21.6 \pm 2.8$ \\
$8166.54872$ & $+32.0 \pm 2.8$ \\
$8168.34173$ & $+45.2 \pm 2.8$ \\
$8170.43537$ & $+54.0 \pm 2.8$ \\
$8171.40315$ & $+59.2 \pm 2.8$ \\
$8172.49950$ & $+55.8 \pm 2.8$ \\
$8173.52487$ & $+54.3 \pm 2.8$ \\
$8174.48287$ & $+56.7 \pm 2.8$ \\
$8175.37457$ & $+43.4 \pm 2.8$ \\
$8176.46248$ & $+40.2 \pm 2.8$ \\
$8177.31903$ & $+40.7 \pm 2.8$ \\
$8179.35163$ & $+25.7 \pm 2.8$ \\
$8183.37853$ & $-2.3 \pm 2.8$ \\
$8190.43942$ & $-16.2 \pm 2.8$ \\
$8197.41341$ & $+12.2 \pm 2.8$ \\
$8202.34623$ & $+44.2 \pm 2.8$ \\
$8207.41584$ & $+56.6 \pm 2.8$ \\
$8213.43115$ & $+27.2 \pm 2.8$ \\
$8214.45061$ & $+20.7 \pm 2.8$ \\
$8215.29403$ & $+10.8 \pm 2.8$ \\
$8216.40131$ & $+7.4 \pm 2.8$ \\
$8217.32291$ & $+1.9 \pm 2.8$ \\
$8218.39959$ & $-5.5 \pm 2.8$ \\
$8219.36300$ & $-12.2 \pm 2.8$ \\
$8226.32948$ & $-12.2 \pm 2.8$ \\
$8227.32661$ & $-11.3 \pm 2.8$ \\
$8228.34543$ & $-8.0 \pm 2.8$ \\
$8229.37011$ & $-5.0 \pm 2.8$ \\
$8230.34292$ & $-1.3 \pm 2.8$ \\
$8236.35027$ & $+32.5 \pm 2.8$ \\
$8238.34104$ & $+46.9 \pm 2.8$ \\
$8240.35399$ & $+46.2 \pm 2.8$ \\
$8243.32639$ & $+50.8 \pm 2.8$ \\
$8244.33485$ & $+48.9 \pm 2.8$ \\
$8245.33729$ & $+37.9 \pm 2.8$ \\
$8246.33954$ & $+35.8 \pm 2.8$ \\
$8252.32785$ & $+2.3 \pm 2.8$ \\
$8253.32727$ & $-4.2 \pm 2.8$ \\
\end{supertabular}
\end{center}

\newpage
\begin{center}
\vspace{0.8cm}
 \tablefirsthead{%
\hline
Date of Obs.         & \multicolumn{1}{c}{RV} \\
		$\text{BJD}-2450000$ & \multicolumn{1}{c}{$[$km/s$]$} \\
		\hline}
\tablehead{%
\multicolumn{1}{l}{Continued}\\
\hline
Date of Obs.         & \multicolumn{1}{c}{RV} \\
		$\text{BJD}-2450000$ & \multicolumn{1}{c}{$[$km/s$]$} \\
		\hline}
\tabletail{%
\hline	}
\tablelasttail{\hline}
\tablecaption{RV measurements of HIP\,85829}
\begin{supertabular}{c @{\hspace{2cm}} r}
$7728.28416$ & $-22.8 \pm 0.8$\\
$7729.21441$ & $-23.5 \pm 0.8$\\
$7743.23837$ & $-21.2 \pm 0.8$\\
$7744.18086$ & $-20.7 \pm 0.8$\\
$7759.60248$ & $-11.1 \pm 0.8$\\
$7760.19004$ & $-12.0 \pm 0.8$\\
$7764.24887$ & $-19.8 \pm 0.8$\\
$7773.62955$ & $-27.0 \pm 0.8$\\
$7775.57755$ & $-28.3 \pm 0.8$\\
$7776.59693$ & $-26.2 \pm 0.8$\\
$7780.57779$ & $-23.6 \pm 0.8$\\
$7782.59454$ & $-19.7 \pm 0.8$\\
$7798.67166$ & $-13.3 \pm 0.8$\\
$7799.65676$ & $-15.0 \pm 0.8$\\
$7800.65083$ & $-16.6 \pm 0.8$\\
$7826.44009$ & $-7.9 \pm 0.8$\\
$7840.38590$ & $-19.2 \pm 0.8$\\
$7841.34600$ & $-21.9 \pm 0.8$\\
$7843.38080$ & $-24.2 \pm 0.8$\\
$7864.37753$ & $-8.0 \pm 0.8$\\
$7874.36547$ & $-11.6 \pm 0.8$\\
$7876.39810$ & $-15.1 \pm 0.8$\\
$7880.38006$ & $-21.7 \pm 0.8$\\
$8151.62384$ & $-26.6 \pm 0.8$\\
$8154.54982$ & $-28.0 \pm 0.8$\\
$8155.63042$ & $-24.6 \pm 0.8$\\
$8158.71864$ & $-23.7 \pm 0.8$\\
$8163.57372$ & $-16.9 \pm 0.8$\\
$8164.59799$ & $-15.4 \pm 0.8$\\
$8166.52981$ & $-12.2 \pm 0.8$\\
$8166.62025$ & $-12.1 \pm 0.8$\\
$8170.48567$ & $-5.8 \pm 0.8$\\
$8171.48526$ & $-5.3 \pm 0.8$\\
$8172.54915$ & $-4.9 \pm 0.8$\\
$8173.55009$ & $-5.4 \pm 0.8$\\
$8174.63598$ & $-5.4 \pm 0.8$\\
$8177.51064$ & $-9.4 \pm 0.8$\\
$8178.50990$ & $-11.3 \pm 0.8$\\
$8179.52909$ & $-13.2 \pm 0.8$\\
$8190.48728$ & $-25.5 \pm 0.8$ \\
$8197.47083$ & $-23.4 \pm 0.8$ \\
$8202.49429$ & $-15.1 \pm 0.8$ \\
$8207.63888$ & $-7.7 \pm 0.8$ \\
$8213.50249$ & $-7.5 \pm 0.8$ \\
$8214.63780$ & $-8.6 \pm 0.8$ \\
$8215.52753$ & $-9.2 \pm 0.8$ \\
$8216.58140$ & $-11.7 \pm 0.8$ \\
$8217.58282$ & $-14.8 \pm 0.8$ \\
$8218.43050$ & $-15.3 \pm 0.8$ \\
$8219.41335$ & $-16.7 \pm 0.8$ \\
$8220.49339$ & $-17.5 \pm 0.8$ \\
$8223.45411$ & $-23.6 \pm 0.8$ \\
$8226.39979$ & $-24.0 \pm 0.8$ \\
$8227.41248$ & $-25.1 \pm 0.8$ \\
$8228.39390$ & $-26.4 \pm 0.8$ \\
$8229.39840$ & $-27.2 \pm 0.8$ \\
$8230.35739$ & $-26.7 \pm 0.8$ \\
$8232.36948$ & $-26.6 \pm 0.8$ \\
$8236.53855$ & $-24.4 \pm 0.8$ \\
$8238.39563$ & $-20.9 \pm 0.8$ \\
$8239.44895$ & $-17.6 \pm 0.8$ \\
$8240.40113$ & $-18.3 \pm 0.8$ \\
$8243.38329$ & $-12.3 \pm 0.8$ \\
$8244.38503$ & $-11.0 \pm 0.8$ \\
$8245.38966$ & $-7.1 \pm 0.8$ \\
$8246.40243$ & $-5.9 \pm 0.8$ \\
$8248.42605$ & $-5.3 \pm 0.8$ \\
$8252.39887$ & $-8.4 \pm 0.8$ \\
$8256.39567$ & $-14.4 \pm 0.8$ \\
$8259.37366$ & $-19.9 \pm 0.8$ \\
$8260.37372$ & $-21.1 \pm 0.8$ \\
$8264.40299$ & $-24.0 \pm 0.8$ \\
$8265.39885$ & $-26.1 \pm 0.8$ \\
$8267.39249$ & $-27.5 \pm 0.8$ \\
$8269.38291$ & $-26.4 \pm 0.8$ \\
$8272.38151$ & $-26.8 \pm 0.8$ \\
$8273.36242$ & $-24.3 \pm 0.8$ \\
\end{supertabular}
\end{center}

\begin{center}
\vspace{0.8cm}
 \tablefirsthead{%
\hline
Date of Obs.         & \multicolumn{1}{c}{RV} \\
		$\text{BJD}-2450000$ & \multicolumn{1}{c}{$[$km/s$]$} \\
		\hline}
\tablehead{%
\multicolumn{1}{l}{Continued}\\
\hline
Date of Obs.         & \multicolumn{1}{c}{RV} \\
		$\text{BJD}-2450000$ & \multicolumn{1}{c}{$[$km/s$]$} \\
		\hline}
\tabletail{%
\hline	}
\tablelasttail{\hline}
\tablecaption{RV measurements of HIP\,77986}
\begin{supertabular}{c @{\hspace{2cm}} r}
$7728.72170$ & $-12.5 \pm 2.6$ \\
$7743.65296$ & $-20.1 \pm 2.6$ \\
$7744.65014$ & $-17.0 \pm 2.6$ \\
$7759.59463$ & $-29.9 \pm 2.6$ \\
$7773.62136$ & $-13.4 \pm 2.6$ \\
$7775.56958$ & $-9.8 \pm 2.6$ \\
$7776.57561$ & $-9.0 \pm 2.6$ \\
$7780.57046$ & $-7.7 \pm 2.6$ \\
$7782.58725$ & $-12.4 \pm 2.6$ \\
$7798.66521$ & $-19.3 \pm 2.6$ \\
$7799.65017$ & $-23.8 \pm 2.6$ \\
$7800.64464$ & $-24.3 \pm 2.6$ \\
$7826.43460$ & $-12.2 \pm 2.6$ \\
$7840.38068$ & $-16.5 \pm 2.6$ \\
$7843.37367$ & $-17.1 \pm 2.6$ \\
$7849.45919$ & $-22.8 \pm 2.6$ \\
$7864.37104$ & $-11.7 \pm 2.6$ \\
$7870.38671$ & $-9.1 \pm 2.6$ \\
$7874.35860$ & $-11.8 \pm 2.6$ \\
$7876.38293$ & $-13.3 \pm 2.6$ \\
$7880.37228$ & $-12.1 \pm 2.6$ \\
$8151.61362$ & $-15.1 \pm 2.6$ \\
$8155.61993$ & $-17.3 \pm 2.6$ \\
$8158.64372$ & $-21.5 \pm 2.6$ \\
$8163.56405$ & $-21.6 \pm 2.6$ \\
$8164.58841$ & $-24.5 \pm 2.6$ \\
$8166.51779$ & $-21.4 \pm 2.6$ \\
$8170.47201$ & $-29.5 \pm 2.6$ \\
$8171.49803$ & $-27.4 \pm 2.6$ \\
$8172.53880$ & $-26.9 \pm 2.6$ \\
$8173.53814$ & $-24.0 \pm 2.6$ \\
$8174.62605$ & $-26.7 \pm 2.6$ \\
$8177.49942$ & $-24.9 \pm 2.6$ \\
$8178.49972$ & $-25.8 \pm 2.6$ \\
$8179.51861$ & $-24.0 \pm 2.6$ \\
$8190.45404$ & $-12.7 \pm 2.6$ \\
$8197.46118$ & $-16.0 \pm 2.6$ \\
$8202.40438$ & $-15.4 \pm 2.6$ \\
$8213.45481$ & $-24.8 \pm 2.6$ \\
$8214.58007$ & $-24.5 \pm 2.6$ \\
$8215.51849$ & $-24.7 \pm 2.6$ \\
$8216.57255$ & $-26.0 \pm 2.6$ \\
$8217.57393$ & $-24.9 \pm 2.6$ \\
$8218.44600$ & $-30.7 \pm 2.6$ \\
$8219.40032$ & $-34.6 \pm 2.6$ \\
$8220.48412$ & $-27.0 \pm 2.6$ \\
$8223.43087$ & $-24.9 \pm 2.6$ \\
$8226.39099$ & $-25.0 \pm 2.6$ \\
$8227.40056$ & $-23.5 \pm 2.6$ \\
$8228.37578$ & $-19.2 \pm 2.6$ \\
$8229.38866$ & $-24.1 \pm 2.6$ \\
$8230.36857$ & $-16.7 \pm 2.6$ \\
$8232.38070$ & $-19.1 \pm 2.6$ \\
$8236.52674$ & $-14.9 \pm 2.6$ \\
$8238.38652$ & $-16.8 \pm 2.6$ \\
$8243.37388$ & $-12.9 \pm 2.6$ \\
$8244.37613$ & $-13.7 \pm 2.6$ \\
$8245.38068$ & $-17.8 \pm 2.6$ \\
$8246.39029$ & $-17.0 \pm 2.6$ \\
$8252.38993$ & $-20.0 \pm 2.6$ \\
$8258.37459$ & $-26.0 \pm 2.6$ \\
$8259.36406$ & $-28.9 \pm 2.6$ \\
$8264.39321$ & $-29.7 \pm 2.6$ \\
\end{supertabular}
\end{center}

\begin{center}
\vspace{0.8cm}
 \tablefirsthead{%
\hline
Date of Obs.         & \multicolumn{1}{c}{RV} \\
		$\text{BJD}-2450000$ & \multicolumn{1}{c}{$[$km/s$]$} \\
		\hline}
\tablehead{%
\multicolumn{1}{l}{Continued}\\
\hline
Date of Obs.         & \multicolumn{1}{c}{RV} \\
		$\text{BJD}-2450000$ & \multicolumn{1}{c}{$[$km/s$]$} \\
		\hline}
\tabletail{%
\hline	}
\tablelasttail{\hline}
\tablecaption{RV measurements of HIP\,98194}
\begin{supertabular}{c @{\hspace{2cm}} r}
$7728.33086$ & $-54.6 \pm 2.0$ \\
$7729.26383$ & $-58.0 \pm 2.0$ \\
$7743.24548$ & $+24.4 \pm 2.0$ \\
$7744.18793$ & $+28.6 \pm 2.0$ \\
$7760.19605$ & $+2.0 \pm 2.0$ \\
$7764.25453$ & $-7.3 \pm 2.0$ \\
$7773.19967$ & $-30.7 \pm 2.0$ \\
$7776.20761$ & $-34.0 \pm 2.0$ \\
$7780.21947$ & $-43.7 \pm 2.0$ \\
$7782.21305$ & $-43.1 \pm 2.0$ \\
$7798.67700$ & $-60.4 \pm 2.0$ \\
$7799.66201$ & $-58.2 \pm 2.0$ \\
$7800.66233$ & $-56.5 \pm 2.0$ \\
$7840.51943$ & $-22.2 \pm 2.0$ \\
$7841.51707$ & $-23.3 \pm 2.0$ \\
$7843.51749$ & $-28.4 \pm 2.0$ \\
$7849.53985$ & $-38.7 \pm 2.0$ \\
$7874.42166$ & $-45.8 \pm 2.0$ \\
$7876.46334$ & $-34.9 \pm 2.0$ \\
$7880.49384$ & $+0.8 \pm 2.0$ \\
$8151.72436$ & $-56.2 \pm 2.0$ \\
$8155.71629$ & $-43.1 \pm 2.0$ \\
$8163.72153$ & $+20.2 \pm 2.0$ \\
$8171.62854$ & $+31.5 \pm 2.0$ \\
$8172.60912$ & $+25.5 \pm 2.0$ \\
$8173.62579$ & $+25.0 \pm 2.0$ \\
$8174.64352$ & $+22.1 \pm 2.0$ \\
$8177.62698$ & $+13.1 \pm 2.0$ \\
$8178.64245$ & $+11.7 \pm 2.0$ \\
$8179.64848$ & $+4.5 \pm 2.0$ \\
$8197.60999$ & $-33.8 \pm 2.0$ \\
$8202.53485$ & $-42.4 \pm 2.0$ \\
$8207.53309$ & $-52.7 \pm 2.0$ \\
$8213.57149$ & $-59.9 \pm 2.0$ \\
$8215.53530$ & $-62.9 \pm 2.0$ \\
$8216.58840$ & $-62.0 \pm 2.0$ \\
$8217.58957$ & $-65.3 \pm 2.0$ \\
$8220.50220$ & $-62.9 \pm 2.0$ \\
$8223.46652$ & $-60.3 \pm 2.0$ \\
$8227.52998$ & $-31.5 \pm 2.0$ \\
$8228.53513$ & $-26.7 \pm 2.0$ \\
$8229.52736$ & $-15.8 \pm 2.0$ \\
$8230.46303$ & $-10.4 \pm 2.0$ \\
$8236.54768$ & $+30.7 \pm 2.0$ \\
$8240.49987$ & $+31.5 \pm 2.0$ \\
$8243.43850$ & $+26.7 \pm 2.0$ \\
$8245.44952$ & $+21.6 \pm 2.0$ \\
$8246.45394$ & $+14.4 \pm 2.0$ \\
$8248.59410$ & $+11.2 \pm 2.0$ \\
$8252.40786$ & $+1.9 \pm 2.0$ \\
$8256.40448$ & $-11.3 \pm 2.0$ \\
$8259.38326$ & $-13.5 \pm 2.0$ \\
$8260.38323$ & $-16.9 \pm 2.0$ \\
$8264.41324$ & $-30.2 \pm 2.0$ \\
$8265.41290$ & $-26.8 \pm 2.0$ \\
$8267.40148$ & $-34.9 \pm 2.0$ \\
$8269.39227$ & $-41.5 \pm 2.0$ \\
$8272.41428$ & $-45.0 \pm 2.0$ \\
$8273.39399$ & $-45.8 \pm 2.0$ \\
\end{supertabular}
\end{center}

\begin{center}
\vspace{0.8cm}
 \tablefirsthead{%
\hline
Date of Obs.         & \multicolumn{1}{c}{RV} \\
		$\text{BJD}-2450000$ & \multicolumn{1}{c}{$[$km/s$]$} \\
		\hline}
\tablehead{%
\multicolumn{1}{l}{Continued}\\
\hline
Date of Obs.         & \multicolumn{1}{c}{RV} \\
		$\text{BJD}-2450000$ & \multicolumn{1}{c}{$[$km/s$]$} \\
		\hline}
\tabletail{%
\hline	}
\tablelasttail{\hline}
\tablecaption{RV measurements of HIP\,107136}
\begin{supertabular}{c @{\hspace{2cm}} r}
$7728.34768$ & $+0.8 \pm 3.1$ \\
$7729.27877$ & $-1.2 \pm 3.1$ \\
$7743.25461$ & $-5.6 \pm 3.1$ \\
$7744.21207$ & $-0.6 \pm 3.1$ \\
$7760.23788$ & $-5.6 \pm 3.1$ \\
$7764.26354$ & $-2.5 \pm 3.1$ \\
$7773.21582$ & $-7.1 \pm 3.1$ \\
$7775.27287$ & $-4.7 \pm 3.1$ \\
$7776.23554$ & $-5.9 \pm 3.1$ \\
$7782.23981$ & $-3.4 \pm 3.1$ \\
$7798.27627$ & $+0.7 \pm 3.1$ \\
$7799.28929$ & $-2.8 \pm 3.1$ \\
$7800.25683$ & $-3.2 \pm 3.1$ \\
$7840.52799$ & $-9.2 \pm 3.1$ \\
$7841.52315$ & $-7.7 \pm 3.1$ \\
$7843.60106$ & $-0.5 \pm 3.1$ \\
$7874.45321$ & $-4.4 \pm 3.1$ \\
$7880.44362$ & $-1.5 \pm 3.1$ \\
$8146.24573$ & $-8.2 \pm 3.1$ \\
$8151.22587$ & $-7.6 \pm 3.1$ \\
$8155.29609$ & $-6.7 \pm 3.1$ \\
$8162.31356$ & $-9.0 \pm 3.1$ \\
$8171.66083$ & $-6.7 \pm 3.1$ \\
$8172.65879$ & $-4.7 \pm 3.1$ \\
$8173.64684$ & $-5.6 \pm 3.1$ \\
$8174.68695$ & $-2.1 \pm 3.1$ \\
$8176.63254$ & $-3.6 \pm 3.1$ \\
$8178.65356$ & $-5.1 \pm 3.1$ \\
$8197.62064$ & $-7.7 \pm 3.1$ \\
$8202.57024$ & $-2.5 \pm 3.1$ \\
$8207.54446$ & $-6.3 \pm 3.1$ \\
$8214.64218$ & $-8.0 \pm 3.1$ \\
$8215.60049$ & $-5.2 \pm 3.1$ \\
$8216.65166$ & $-6.3 \pm 3.1$ \\
$8217.64715$ & $-7.7 \pm 3.1$ \\
$8219.57666$ & $-10.6 \pm 3.1$ \\
$8220.59183$ & $-10.6 \pm 3.1$ \\
$8228.59445$ & $-7.8 \pm 3.1$ \\
$8229.59686$ & $-10.6 \pm 3.1$ \\
$8230.56516$ & $-11.1 \pm 3.1$ \\
$8232.55260$ & $-4.9 \pm 3.1$ \\
$8238.46692$ & $-9.6 \pm 3.1$ \\
$8240.53233$ & $-9.5 \pm 3.1$ \\
$8243.52378$ & $-3.8 \pm 3.1$ \\
$8244.52652$ & $-7.7 \pm 3.1$ \\
$8245.50747$ & $-3.0 \pm 3.1$ \\
$8246.51472$ & $-5.3 \pm 3.1$ \\
$8252.58164$ & $-6.7 \pm 3.1$ \\
$8256.41395$ & $-3.3 \pm 3.1$ \\
$8264.47936$ & $-8.4 \pm 3.1$ \\
$8265.48615$ & $-5.4 \pm 3.1$ \\
$8267.48187$ & $-7.9 \pm 3.1$ \\
$8269.47253$ & $-10.2 \pm 3.1$ \\
$8272.47911$ & $-4.4 \pm 3.1$ \\
$8273.42054$ & $-11.7 \pm 3.1$ \\
\end{supertabular}
\end{center}

\begin{center}
\vspace{0.8cm}
  \tablefirsthead{%
\hline
Date of Obs.         & \multicolumn{1}{c}{RV} \\
		$\text{BJD}-2450000$ & \multicolumn{1}{c}{$[$km/s$]$} \\
		\hline}
\tablehead{%
\multicolumn{1}{l}{Continued}\\
\hline
Date of Obs.         & \multicolumn{1}{c}{RV} \\
		$\text{BJD}-2450000$ & \multicolumn{1}{c}{$[$km/s$]$} \\
		\hline}
\tabletail{%
\hline	}
\tablelasttail{\hline}
\tablecaption{RV measurements of HIP\,107533}
\begin{supertabular}{c @{\hspace{2cm}} r}
$7728.35402$ & $-16.6 \pm 1.2$ \\
$7729.28583$ & $-15.4 \pm 1.2$ \\
$7743.26866$ & $-16.9 \pm 1.2$ \\
$7744.21863$ & $-16.2 \pm 1.2$ \\
$7760.28400$ & $-17.4 \pm 1.2$ \\
$7764.28035$ & $-16.7 \pm 1.2$ \\
$7773.23083$ & $-15.7 \pm 1.2$ \\
$7775.27993$ & $-16.8 \pm 1.2$ \\
$7776.24328$ & $-16.8 \pm 1.2$ \\
$7782.24702$ & $-17.7 \pm 1.2$ \\
$7798.28327$ & $-16.5 \pm 1.2$ \\
$7799.29815$ & $-17.5 \pm 1.2$ \\
$7800.24873$ & $-16.6 \pm 1.2$ \\
$7840.53464$ & $-15.8 \pm 1.2$ \\
$7841.52975$ & $-14.4 \pm 1.2$ \\
$7843.61561$ & $-17.4 \pm 1.2$ \\
$7874.46144$ & $-17.1 \pm 1.2$ \\
$7880.45123$ & $-15.7 \pm 1.2$ \\
$8146.25345$ & $-16.4 \pm 1.2$ \\
$8151.23414$ & $-18.9 \pm 1.2$ \\
$8155.30427$ & $-17.4 \pm 1.2$ \\
$8162.32888$ & $-16.5 \pm 1.2$ \\
$8171.66904$ & $-17.0 \pm 1.2$ \\
$8172.66630$ & $-16.8 \pm 1.2$ \\
$8173.66925$ & $-18.0 \pm 1.2$ \\
$8174.70697$ & $-18.2 \pm 1.2$ \\
$8176.63996$ & $-18.2 \pm 1.2$ \\
$8178.66027$ & $-17.4 \pm 1.2$ \\
$8179.63646$ & $-17.6 \pm 1.2$ \\
$8197.62828$ & $-18.0 \pm 1.2$ \\
$8202.58052$ & $-18.6 \pm 1.2$ \\
$8207.57900$ & $-17.9 \pm 1.2$ \\
$8215.62021$ & $-16.7 \pm 1.2$ \\
$8216.65828$ & $-16.5 \pm 1.2$ \\
$8217.65830$ & $-16.6 \pm 1.2$ \\
$8219.59906$ & $-16.5 \pm 1.2$ \\
$8220.61231$ & $-18.4 \pm 1.2$ \\
$8228.62179$ & $-17.9 \pm 1.2$ \\
$8230.58527$ & $-18.2 \pm 1.2$ \\
$8238.47419$ & $-15.7 \pm 1.2$ \\
$8240.55271$ & $-16.0 \pm 1.2$ \\
$8243.54366$ & $-16.7 \pm 1.2$ \\
$8244.54650$ & $-18.9 \pm 1.2$ \\
$8245.52753$ & $-17.2 \pm 1.2$ \\
$8246.59662$ & $-18.4 \pm 1.2$ \\
$8252.58853$ & $-15.9 \pm 1.2$ \\
$8264.50000$ & $-18.3 \pm 1.2$ \\
$8265.55263$ & $-16.2 \pm 1.2$ \\
$8267.50238$ & $-19.4 \pm 1.2$ \\
$8269.49318$ & $-18.7 \pm 1.2$ \\
$8272.50371$ & $-20.4 \pm 1.2$ \\
$8273.49114$ & $-19.5 \pm 1.2$ \\
\end{supertabular}
\label{tab:HIP107533}
\end{center}


\begin{thebibliography}{}
\bibitem[Abt et al. 1985]{abt} Abt, H.~A., \& Levy, S.~G. \ 1985, ApJS, 59, 229
\bibitem[Allende Prieto et al. 1999]{allende} Allende Prieto C., Lambert D.~L. \ 1999, A\&A, 352, 555
\bibitem[Batten et al. 1982]{batten} Batten, A.~H., Fisher, W.~A. \& Fletcher, J.~M. \ 1982, PASP, 94, 515
\bibitem[Bischoff et al. 2017]{bischoff} Bischoff, R., Mugrauer, M. Zehe, T. et al. \ 2017, AN, 338, 671
\bibitem[Blaauw et al. 1963]{blaauw} Blaauw, A. \& van Albada, T.~S. \ 1963, ApJ, 137, 791
\bibitem[Dworetsky 1983]{dworetsky} Dworetsky, M.~M., \ 1983, MNRAS, 203, 917
\bibitem[Fehrenbach 1948]{fehrenbach} Fehrenbach, C. \ 1948, AnAp, 11, 157
\bibitem[Gaia Collaboration et al. 2018]{gaia} Gaia Collaboration, et al., 2018, A\&A, 616, A1
\bibitem[Harper 1925]{harper} Harper, W.~E. \ 1925, PDAO, 3, 189
\bibitem[Horch et al. 2015] {horch} Horch, E.~P. et al. \ 2015, AJ, 150, 151
\bibitem[Heard et al. 1975]{heard} Heard, J.~F., Hurkens, R. \&  Harmanec, P. \ 1975, A\&A, 42, 47
\bibitem[Irrgang et al. 2016]{irrgang} Irrgang, A., Desphande, A. \& Moehler, S.  \ 2016, A\&A, 591, 6
\bibitem[Johnson 2004]{johnson} Johnson, D. O. \ 2004, JAD, 10, 3
\bibitem[Lucy et al. 1971]{lucy} Lucy, L.~B., \& Sweeney, M. A. \ 1971, AJ, 76, 544
\bibitem[Luyten 1936]{luyten}  Luyten, W.~J. \ 1936, ApJ, 84, 85
\bibitem[Mugrauer et al. 2014]{mugrauer2014} Mugrauer, M., Avila, G. \& Guirao, C. \ 2014, AN, 335, 417
\bibitem[Mugrauer et al. 2016]{mugrauer2016} Mugrauer, M., Buder, S., Reum, F. et al. 2016, AN, 337, 226
\bibitem[Ochsenbein, Bauer \& Marcout 2000]{ochsenbein} Ochsenbein F., Bauer P., Marcout J., 2000, A\&AS, 143, 23
\bibitem[Pecaut et al. 2013]{pecaut} Pecaut M.~J. \& Mamajek E.~E. \ 2013, ApJS, 208, 9
\bibitem[Pfau 1984]{pfau} Pfau, W. \ 1984, JenRu, 29, 121
\bibitem[Pourbaix et al. 2004]{pourbaix2004} Pourbaix, D. et al. \ 2004, A\&A, 424, 727
\bibitem[Stickland 1987]{stickland} Stickland, D.~J. \ 1987, Obs, 107, 5
\bibitem[Taffara 1939]{taffara} Taffara, S. \ 1939, MmSAI, 12, 279
\bibitem[Udick 1912]{udick} Udick, S. \ 1912, PAllO, 2, 191
\bibitem[Zehe et al. 2018]{zehe2018} Zehe, T., Mugrauer, M., Neuh\"{a}user, R. et al.\ 2018, AN, 339, 46

\end{thebibliography}
\end{document}